\documentclass[amsmath,amssymb,aps,prd,longbibliography,preprintnumbers,showpacs,floatfix,nofootinbib]
{revtex4-1}

\usepackage{color,graphicx,epsfig} 
 
\def\et{{\rm E}_{\rm T}}
\newcommand\nn{\nonumber}
\newcommand\ba{\begin{eqnarray}}
\newcommand\ea{\end{eqnarray}}

\begin{document}

\preprint{DESY 12-063 \;\;\;\;  FTUAM-2012-10}

\title{Transverse Energy-Energy Correlations in Next-to-Leading Order in $\alpha_s$
at the LHC}

\author{Ahmed~Ali $^a$}
\email{ahmed.ali@desy.de}
\author{ Fernando~Barreiro$^b$}
\email{fernando.barreiro@uam.es}
\author{Javier~Llorente$^b$}
\email{javier.llorente.merino@cern.ch}
\author{Wei~Wang$^a$}
\email{wei.wang@desy.de}
\affiliation{\it a Deutsches Elektronen-Synchrotron DESY, D-22607 Hamburg, Germany}  
\affiliation{\it b Universidad Autonoma de Madrid (UAM), Facultad de Ciencias C-XI, Departamento
de Fisica, Cantoblanco, Madrid, Spain}

\date{\today}

\begin{abstract}
We compute the transverse energy-energy correlation (EEC) and its asymmetry (AEEC) 
in next-to-leading order (NLO) in $\alpha_s$ in proton-proton collisions at the LHC with the 
center-of-mass energy $E_{\rm c.m.}=7$ TeV. We show that the transverse EEC and the AEEC
distributions are insensitive to the QCD
factorization- and the renormalization-scales, structure functions of the proton, and
for a judicious choice of the jet-size, also the underlying minimum bias events.
Hence they can be used to precisely test QCD in hadron colliders and
determine the strong coupling $\alpha_s$. We illustrate these features by defining
the hadron jets using the anti-$k_T$ jet algorithm
and an event selection procedure employed in the analysis of jets at the LHC and show the
$\alpha_s(M_Z)$-dependence of the transverse EEC and the AEEC in the
anticipated range $0.11 \leq \alpha_s(M_Z) \leq 0.13$.

\end{abstract}


\maketitle

%
%
\section{Introduction}
Hadron jets are powerful quantitative tools to study Quantum Chromo Dynamics (QCD)
in high energy physics. In $e^+e^-$  colliders
PETRA, PEP and LEP, and also in  the electron-proton collider HERA, jet studies have been
undertaken extensively. These include the measurements of the inclusive variables, such
as thrust, acoplanarity and hadron energy flow, as well as the exclusive jet distributions,
yielding a consistent and precise value of the
QCD coupling constant $\alpha_s(M_Z)$~\cite{Nakamura:2010zzi} . At the hadron colliders Tevatron and
the LHC, QCD predictions for jets have been compared with the measured transverse momentum ($p_T$)
distributions, and also with the multi-jet
rates~\cite{Wobisch:2012iu,Chatrchyan:2012pb,Aad:2010ad}
assuming a jet algorithm~\cite{Catani:1993hr,Catani:1996vz,Cacciari:2008gp}. The theoretical framework
for calculating the jet cross sections in hadronic collisions in the next-to-leading order (NLO)
accuracy has been in place for well over a decade~\cite{Catani:1996jh,Nagy:2001fj}, which has been
employed in the QCD-based analysis of the data.

 In comparison to the $e^+e^-$ and the $ep$ experiments, event shape variables have so far received
less attention in the analysis of the  data from the hadron colliders, though first results have
been lately published on the measurement of
 the transverse thrust and the thrust minor distributions~\cite{Banfi:2004nk} by the CDF
 collaboration~\cite{Aaltonen:2011et}.  Studies of the hadronic
event shapes in $pp$ collisions at the LHC have just started, initiated
by the CMS collaboration using the central transverse thrust and the central transverse
minor variables, where the term {\it central} refers to the jets in the central
region of the detector~\cite{Khachatryan:2011dx}.
This is followed by a similar analysis by the ATLAS collaboration
~\cite{Aad:2012np}. The distributions in these variables have been
compared with a number of Monte Carlo (MC) simulations, with PYTHIA6~\cite{Sjostrand:2006za}, 
PYTHIA8~\cite{Sjostrand:2007gs} and HERWIG++~\cite{Bahr:2008pv} providing
a good description of the data. However, a bench-mark in this field, namely a quantitative
determination of $\alpha_s(M_Z)$ at the LHC
from the   analysis of data on event shapes, is still very much a work in progress.

In this paper, we calculate the transverse energy energy correlation and its asymmetry proposed
some time ago~\cite{Ali:1984yp} as a quantitative measure of perturbative QCD 
in hadronic collisions. The analogous
energy-energy correlation (EEC) function measurements - the energy weighted angular distributions of
the produced hadron pairs in $e^+e^-$ annihilation - were proposed by
 Basham {et al.}~\cite{Basham:1978bw}. The EEC and
its asymmetry (AEEC) were subsequently calculated in
 $O(\alpha_s^2)$~\cite{Ali:1982ub,Richards:1983sr},
 and their measurements have impacted significantly
on the precision tests of perturbative QCD and in the determination of
$\alpha_s$ in $e^+e^-$ annihilation
 experiments (for a recent review, see~\cite{Ali:2010tw}). 
Transverse EEC distributions in hadronic collisions~\cite{Ali:1984yp}, on the other hand,
are handicapped due to the absence of the  NLO
perturbative QCD corrections.  In the leading order in $\alpha_s(\mu)$, these distributions show
marked sensitivities on the renormalization and factorization scales $\mu=\mu_R$ and
 $\mu=\mu_F$, respectively,
thereby hindering a determination of $\alpha_s(M_Z)$. We aim at remedying this drawback by presenting a
calculation of the transverse EEC  function and its asymmetry in $O(\alpha_s^2(\mu))$,
which reduces the scale-dependence to a few per cent.

The paper is organized as follows.  Sec.~\ref{sec:EEC} collects the definitions and some leading-order features of the transverse energy-energy correlation.  In Sec.~\ref{sec:Results}, we present the numerical results calculated at next-to-leading order  in $\alpha_s$ and demonstrate that the    transverse EEC and its asymmetry  are robust against variations of  various parameters except for $\alpha_s$, for which we present the NLO results in the range $0.11<\alpha_s(m_Z)<0.13$ at the LHC ($\sqrt s =7 $TeV). We conclude in the last section.

%

\section{Transverse Energy-Energy Correlation and its asymmetry} \label{sec:EEC}

We start by recalling the definition of the transverse EEC function~\cite{Ali:1984yp}
\ba
\frac{1}{\sigma^\prime}\frac{d\Sigma^\prime}{d\phi} &\equiv&\frac
{\int_{{\rm E}_{\rm T}^{\rm min}}^{\sqrt{s}}
 d\et\, d^2\Sigma/d\et  \, d\phi}
{\int_{{\rm E}_{\rm T}^{\rm min}}^{\sqrt{s}}\,d\et\, d\sigma/d\et}
 \nn\\
&=& \frac{1}{N}\displaystyle{\sum_{A=1}^{N}} \frac{1}{\Delta \phi}
\displaystyle{\sum_{{\rm pairs~in}~\Delta\phi}}\frac{2{\rm E}_{\rm T a}^{\rm A}\,
{\rm E}_{\rm T b}^{\rm A}}{({\rm E}_{\rm T}^{\rm A})^2}~,
\label{eq:EEC-def}
\ea
with 
\ba
\sigma^\prime = {\int_{{\rm E}_{\rm T}^{\rm min}}^{\sqrt{s}}\,d\et\, d\sigma/d\et}\nn%
\ea

The first sum on the right-hand side in the second of the above equations is over the
events ${\rm A}$ with total transverse energy
 $\et^{\rm A}=\sum_a {\et}_a^{\rm A}\geq \et^{\rm min}$, with the $\et^{\rm min}$ set by
the experimental setup.
 The second sum is over the pairs of
partons ($a,\,b$) whose transverse momenta have relative azimuthal angle $\phi$ to $\phi+\Delta \phi$. 
In addition, the fiducial volume is restricted by the experimental acceptance in the rapidity variable $\eta$. 

In leading order QCD, the transverse energy spectrum $d\sigma/d\et$ is a convolution of the parton distribution
functions (PDFs) with the $2 \to 2$ hard scattering partonic sub-processes.
Away from the end-points, i.e., for $\phi \neq 0^\circ$ and $\phi \neq 180^\circ$, in the leading
order in $\alpha_s$, the energy-weighted cross
section $d^2\Sigma/d\et\,d\phi$ involves the convolution of the PDFs with the $2\to 3$ sub-processes,
such as $gg \to ggg$. Thus, schematically, the leading contribution
for the transverse EEC function is calculated from the following expression:
\begin{equation}
\frac{1}{\sigma^\prime}\frac{d\Sigma^\prime}{d\phi} =
\frac{\Sigma_{a_i,b_i} f_{a_1/p}(x_1) f_{a_2/p}(x_2) \star \hat{\Sigma}^{a_1 a_2 \to b_1 b_2 b_3}}  
     {\Sigma_{a_i,b_i} f_{a_1/p}(x_1) f_{a_2/p}(x_2) \star \hat{\sigma}^{a_1 a_2 \to b_1 b_2 } }~,
\label{eq:EEC-QCD}
\end{equation}
where $\hat{\Sigma}^{a_1 a_2 \to b_1 b_2 b_3}$ is the transverse energy-energy weighted partonic
cross section,
 $x_i$ ($ i=1,2$) are the fractional longitudinal momenta carried by the partons,
$f_{a_1/p}(x_1)$ and $f_{a_2/p}(x_2)$ are the PDFs, and the $\star$ denotes a convolution
 over the appropriate variables.
The function defined in Eq.~(\ref{eq:EEC-QCD}) depends not only on $\phi$, but also on the ratio
$\et^{\rm min}/\sqrt{s}$ and   rapidity   $\eta$. In general, the numerator and the
 denominator in Eq.~(\ref{eq:EEC-QCD})
have a \underline{different} dependence on these variables, as the PDFs are weighted
 differently. However,
as already observed in~\cite{Ali:1984yp}, certain {\it normalized} distributions for the various
sub-processes contributing to the $2 \to 3$ hard scatterings are similar, and the {\it same} combination
of PDFs enters in the $2 \to 2$ and $2 \to 3$ cross sections;  hence  the transverse EEC cross section is
to a good approximation {\it independent} of the PDFs (see, Fig.~1 in~\cite{Ali:1984yp}). Thus,
for a fixed rapidity range $|\eta| < \eta_c$ and the variable $\et/\sqrt{s}$, one has an
approximate factorized result, which in the  LO in $\alpha_s$ reads as 
\begin{equation}
\frac{1}{\sigma^\prime}\frac{d \Sigma^\prime}{d\phi} \sim \frac{\alpha_s(\mu)}{\pi} F(\phi)~,
\label{eq:F-def}
\end{equation}
where 
\begin{eqnarray}
\alpha_s(\mu)= \frac{1}{b_0 \log(\mu^2/\Lambda^2)} \left[1-\frac{b_1 \log(\log(\mu^2/\Lambda^2))}{b_0^2 \log(\mu^2/\Lambda^2)}\right], \;\; b_0= \frac{33-12n_f}{12\pi},\;\; b_1= \frac{153-19n_f}{24\pi^2}. \label{eq:alpha_s}
\end{eqnarray} 
In the above equation, $n_f$ is the active quark flavor number at the scale $\mu$ and the hadronization scale $\Lambda$ is determined by the input $\alpha_s(m_Z)$.  
The function  $F(\phi)$ and the corresponding
 transverse EEC asymmetry defined as 
\begin{equation}
\frac{1}{\sigma^\prime}\frac{d\Sigma^{\prime {\rm asym}}}{d\phi}\equiv
\frac{1}{\sigma^\prime}\frac{d\Sigma^\prime }{d\phi}|_{\phi} -
\frac{1}{\sigma^\prime}\frac{d\Sigma^\prime }{d\phi}|_{\pi -\phi}~,
\end{equation}
were worked out in~\cite{Ali:1984yp} in the leading order of $\alpha_s$ for the CERN SPS $p\bar{p}$
 collider at $\sqrt{s}=540$ GeV.
In particular, it was shown that the transverse EEC functions for the $gg$-, $gq$- and
$q\bar{q}$-scatterings had very similar shapes, and their relative contributions were found
 consistent to a good approximation with the
ratio of the corresponding color factors 1:4/9:16/81 for the $gg$, $gq(=g\bar{q})$ and $q\bar{q}$
 initial  states over a large range of $\phi$.


\section{Next-to-leading order results for the transverse EEC and its asymmetry} \label{sec:Results}
We have used the existing
program NLOJET++~\cite{Nagy:2001fj}, which has been checked in a number of
 independent 
NLO jet calculations~\cite{Bern:2011ep}, to compute the transverse EEC and its asymmetry AEEC in the
 NLO accuracy for the LHC proton-proton center-of-mass energy $\sqrt{s}=7$ TeV. Schematically,
this entails the calculations of the $2 \to 3$ partonic sub-processes in the NLO accuracy and
of the $2 \to 4$  partonic processes in the leading order in $\alpha_s(\mu)$, which contribute
to the numerator on the r.h.s.~of Eq.~(\ref{eq:EEC-QCD}).
We have  restricted the azimuthal
angle range by cutting out regions near $\phi=0^\circ$ and $\phi=180^\circ$.
This would, in particular,
remove the self-correlations ($a=b$) and frees us from calculating the $O(\alpha_s^2)$ (or two-loop)
virtual corrections to the $2 \to 2$ processes. Thus, with the azimuthal angle cut, the numerator
in Eq.~(\ref{eq:EEC-QCD}) is calculated from the $2 \to 3$ and $2 \to 4$ processes to
 $O(\alpha_s^4)$. The denominator in
 Eq.~(\ref{eq:EEC-QCD}) includes the $2 \to 2$  and $2 \to 3$ processes,
which are calculated up to and including the $O(\alpha_s^3)$ corrections.

 In the NLO accuracy, one can express the EEC cross section as  
\begin{equation}
\frac{1}{\sigma^\prime}\frac{d \Sigma^\prime}{d\phi} \sim \frac{\alpha_s(\mu)}{\pi} F(\phi)
\left[1 + \frac{\alpha_s(\mu)}{\pi} G(\phi)\right]~.
\label{eq:G-def}
\end{equation}
It is customary to lump the NLO corrections in a so-called $K$-factor (which,  as shown here, is
a non-trivial function of $\phi$), defined as
\begin{equation}
K^{\rm EEC}(\phi)\equiv 1 + \frac{\alpha_s(\mu)}{\pi} G(\phi) ~. 
\label{eq:K-def}
\end{equation} 
The transverse EEC asymmetry in the NLO accuracy is likewise defined as  
\begin{equation}
\frac{1}{\sigma^\prime}\frac{d \Sigma^{\prime \rm asym}}{d\phi} \sim \frac{\alpha_s(\mu)}{\pi} A(\phi)
\left[1 + \frac{\alpha_s(\mu)}{\pi} B(\phi)\right]~.
\label{eq:AB-def}
\end{equation}
and the corresponding K-factor is defined as 
\begin{equation}
K^{\rm AEEC}(\phi)\equiv 1 + \frac{\alpha_s(\mu)}{\pi} B(\phi) ~. 
\label{eq:K-AEEC-def}
\end{equation} 
The principal result of this paper is the calculation of the NLO functions $K^{\rm EEC}(\phi)$ and $K^{\rm AEEC}(\phi)$ and 
in demonstrating the
insensitivity of the EEC and the AEEC functions, calculated to NLO accuracy, to the various
intrinsic parametric and the underlying event uncertainties.

We now give details of the computations:
In transcribing the NLOJET++~\cite{Nagy:2001fj} program, we have replaced the default
structure functions  therein by the
state of the art PDFs,  for which we use the MSTW~\cite{Martin:2009iq}
and the CT10~\cite{Lai:2010vv} sets. We have also replaced the $k_T$ jet algorithm
by the anti-$k_T$ jet algorithm~\cite{Cacciari:2008gp} for defining the jets, in which the distance measures of partons  are 
given by
\begin{eqnarray}
 d_{ij}= min(k_{ti}^{-2}, k_{tj}^{-2}) \frac{(\eta_i-\eta_j)^2+ (\phi_i-\phi_j)^2}{R^2},\;\;\; d_{iB}= k_{ti}^{-2},
\end{eqnarray}  
with $R$ being the usual radius parameter. 
 We recall that the NLO corrections we are using~\cite{Nagy:2001fj}
have been computed in the Catani-Seymour dipole formalism~\cite{Catani:1996vz}.
In particular, it involves  a certain cutting of the phase space of the dipole subtraction
terms and the numerical calculations require the generation of a very large number of
events (we have generated $O(10^{10})$ events on the DESY-Theory PC Cluster) to bring the statistical
accuracy in the NLO EEC distribution to the desired level of below a few per cent. 
We have assumed the rapidity range $\vert \eta \vert \leq 2.5$, have
put a cut on the transverse energy $E_T>25$ GeV for each jet and  require  
 $E_{T1}+E_{T2}>500$ GeV for the two leading jets.
 The latter  cut ensures that the trigger efficiencies for the LHC detectors will be
 close to $100\%$. We
 have set the transverse energy of the hardest jet  as the default factorization-
 and renormalization-scale , i.e., 
 $\mu_F=\mu_R=  E_{T}^{\rm max}$. We then vary the scales $\mu_F$ and $\mu_R$ independently in the range
 $0.5 \,E_{T}^{\rm max} \leq (\mu_F,\mu_R)
\leq 2  \, E_{T}^{\rm max}$ to study numerically the scale dependence.

\begin{figure}\begin{center}
\includegraphics[scale=0.44]{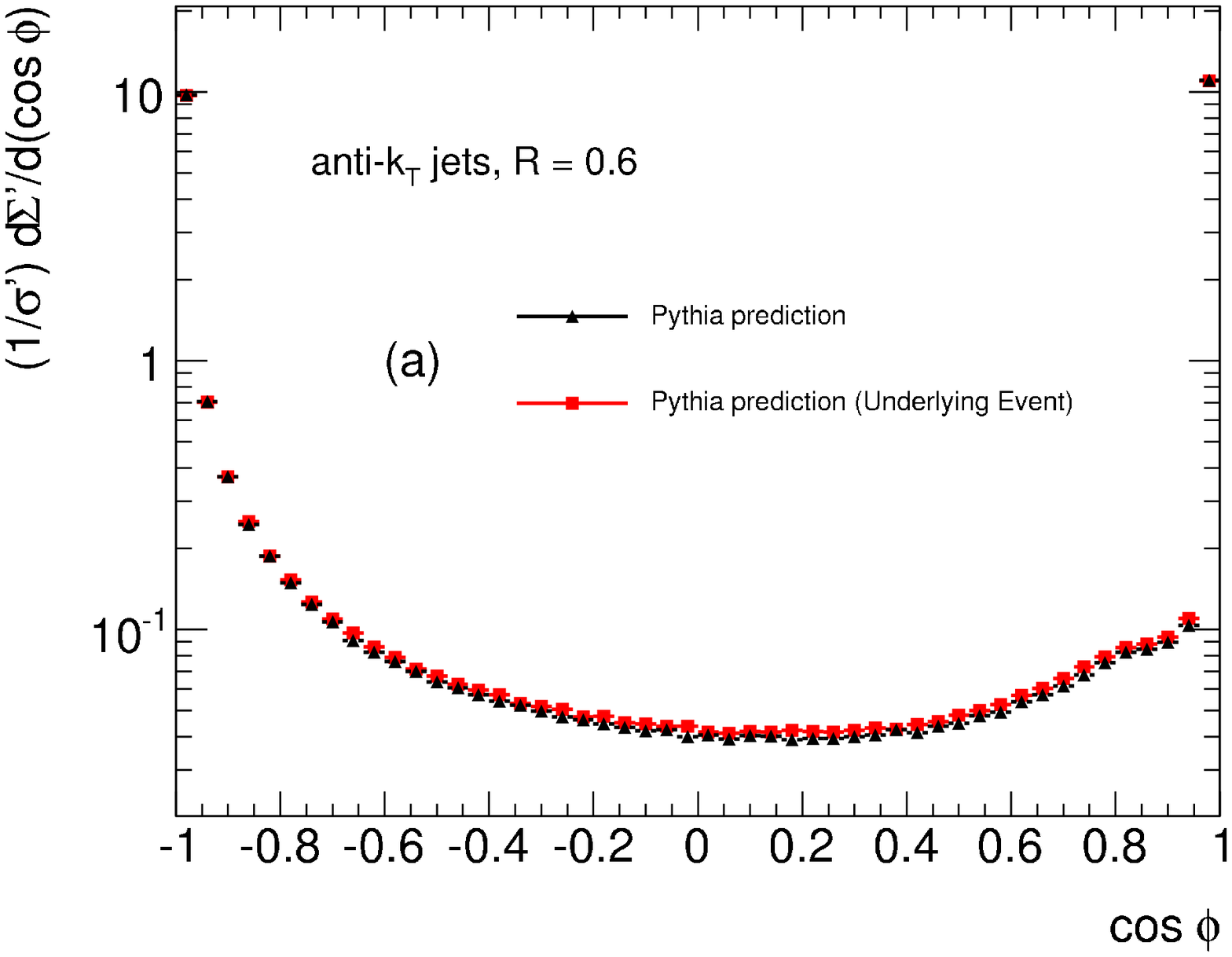}
\includegraphics[scale=0.44]{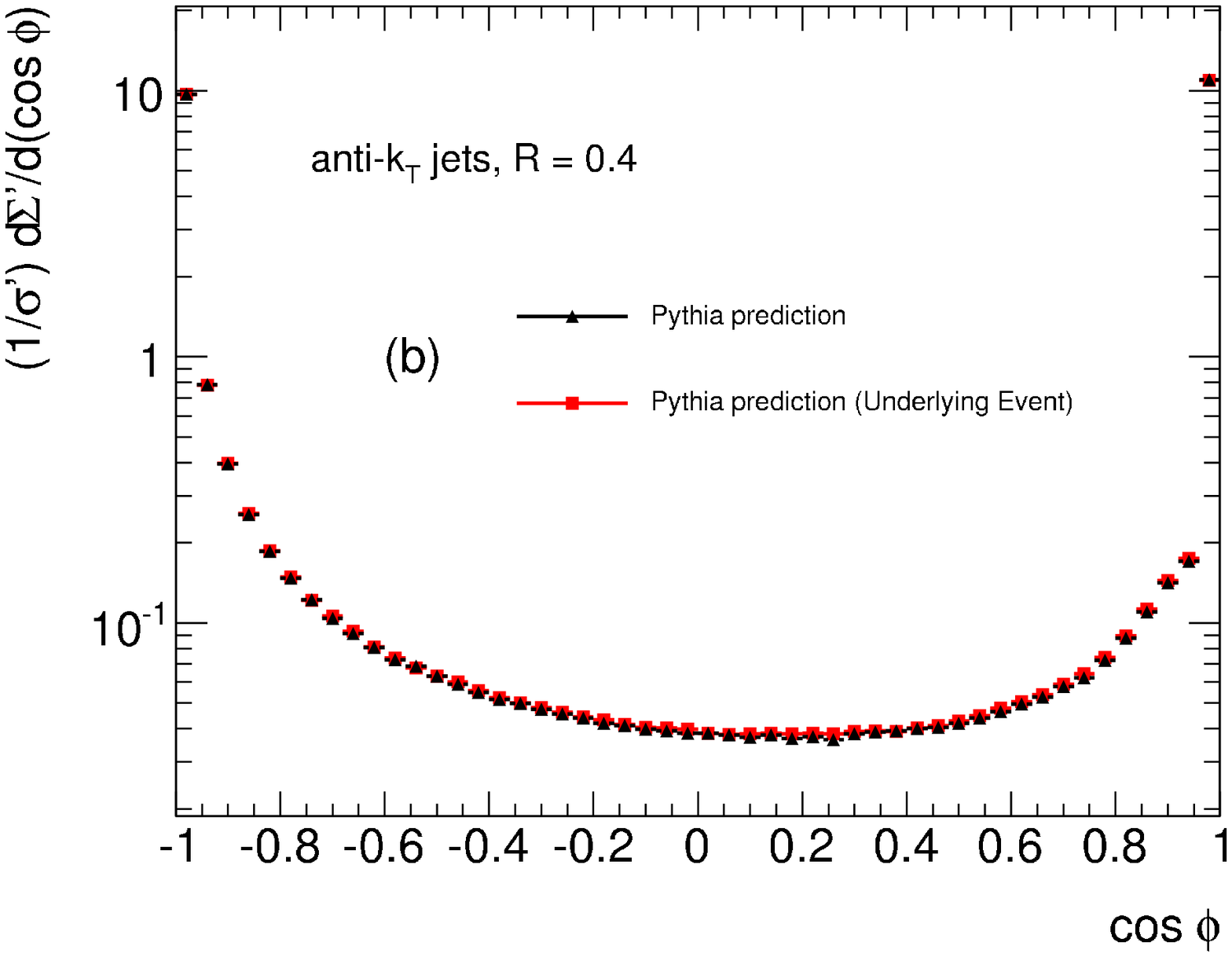} 
\includegraphics[scale=0.44]{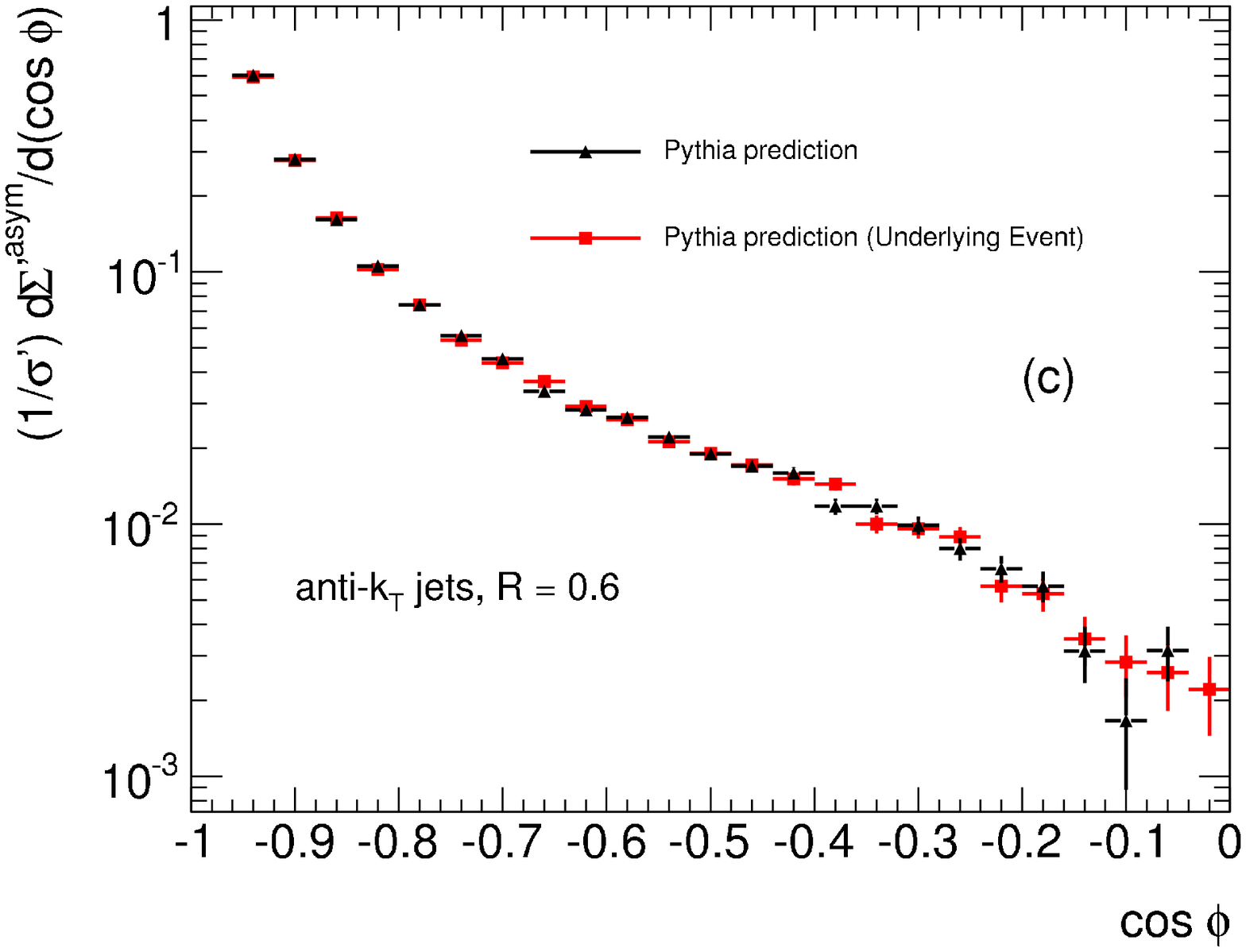}
\includegraphics[scale=0.44]{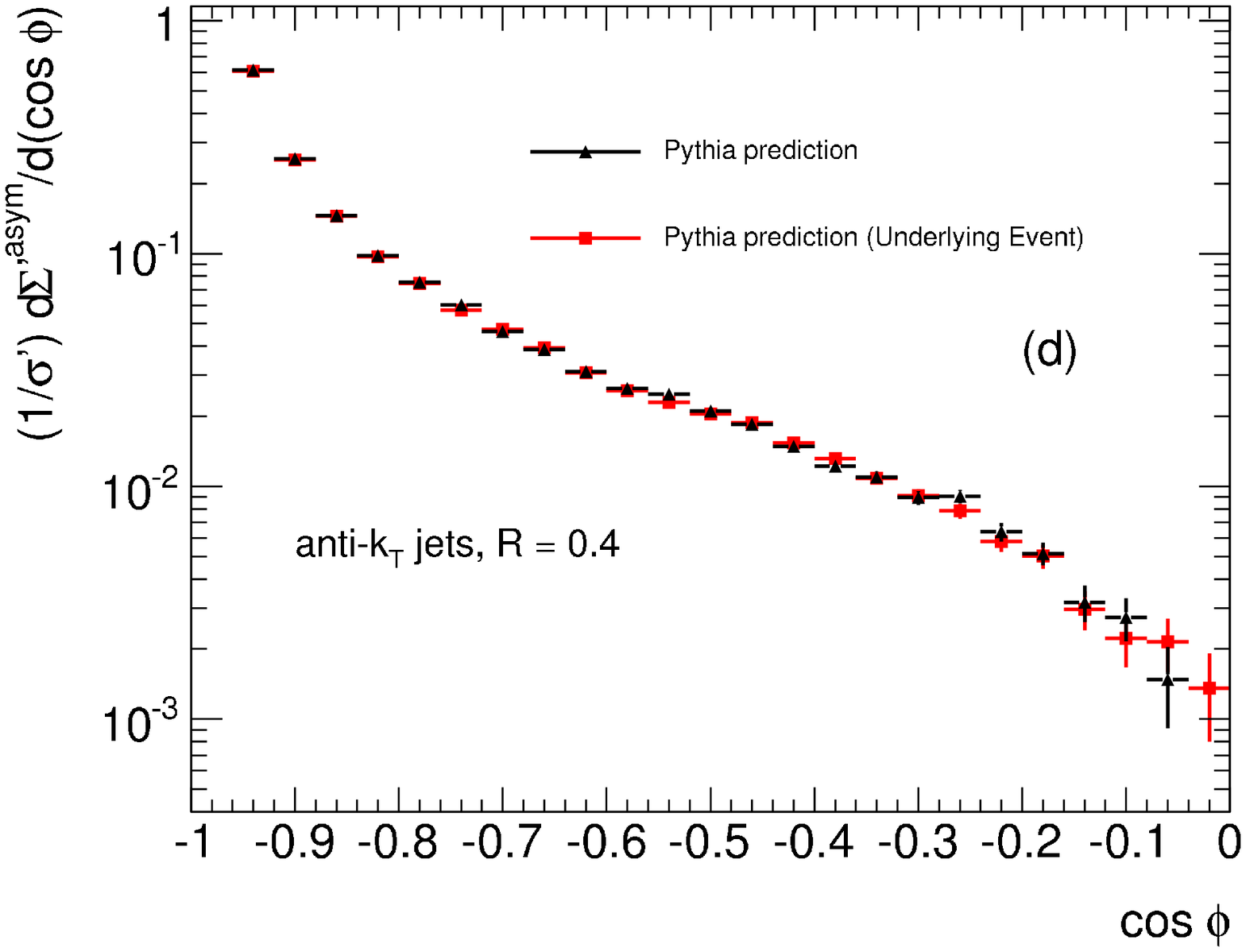} 
\caption{(color online) Differential distribution in $\cos\phi$ of the transverse EEC cross section [(a),(b)] and its asymmetry [(c),(d)] obtained with the  PYTHIA6 MC program~\cite{Sjostrand:2007gs} with and without the underlying events  at $\sqrt s= 7$ TeV and the anti-$k_T$ algorithm with two assumed values of the jet-size parameter $R=0.6$[(a),(c)] and $R=0.4$ [(b),(d)]. } \label{fig:UE}
\end{center}
\end{figure}
%

\begin{figure}\begin{center}
\includegraphics[scale=0.44]{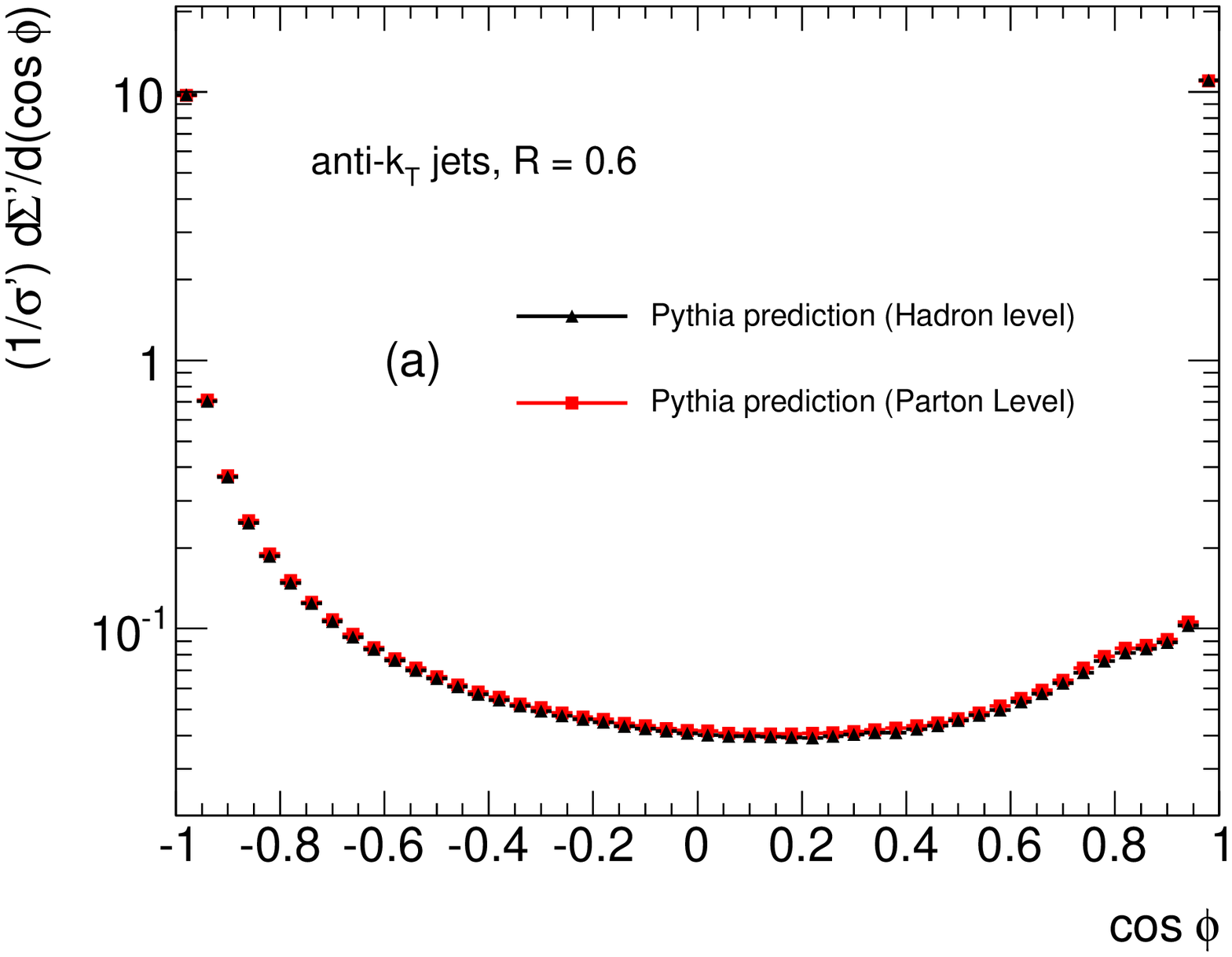}
\includegraphics[scale=0.44]{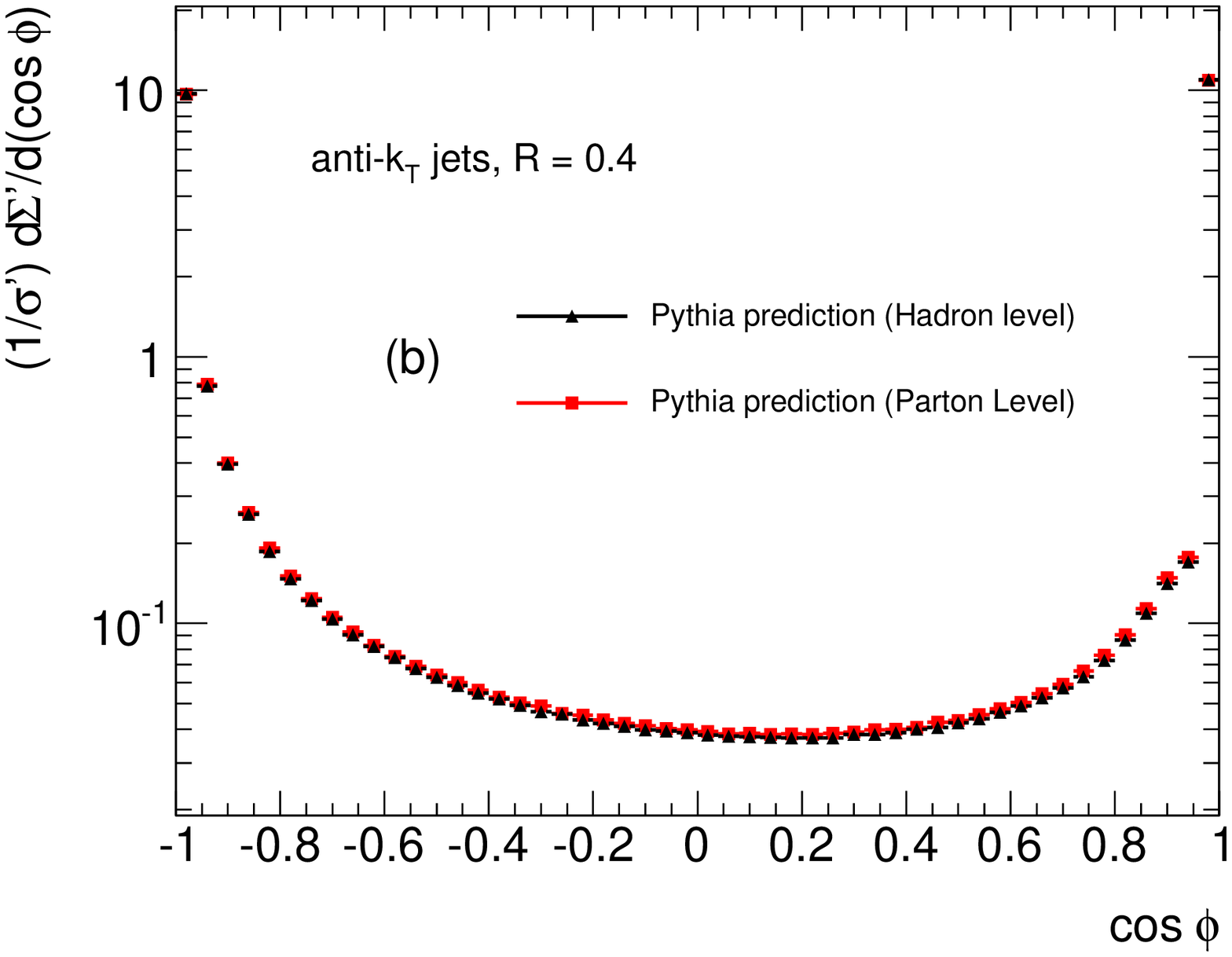} 
\includegraphics[scale=0.44]{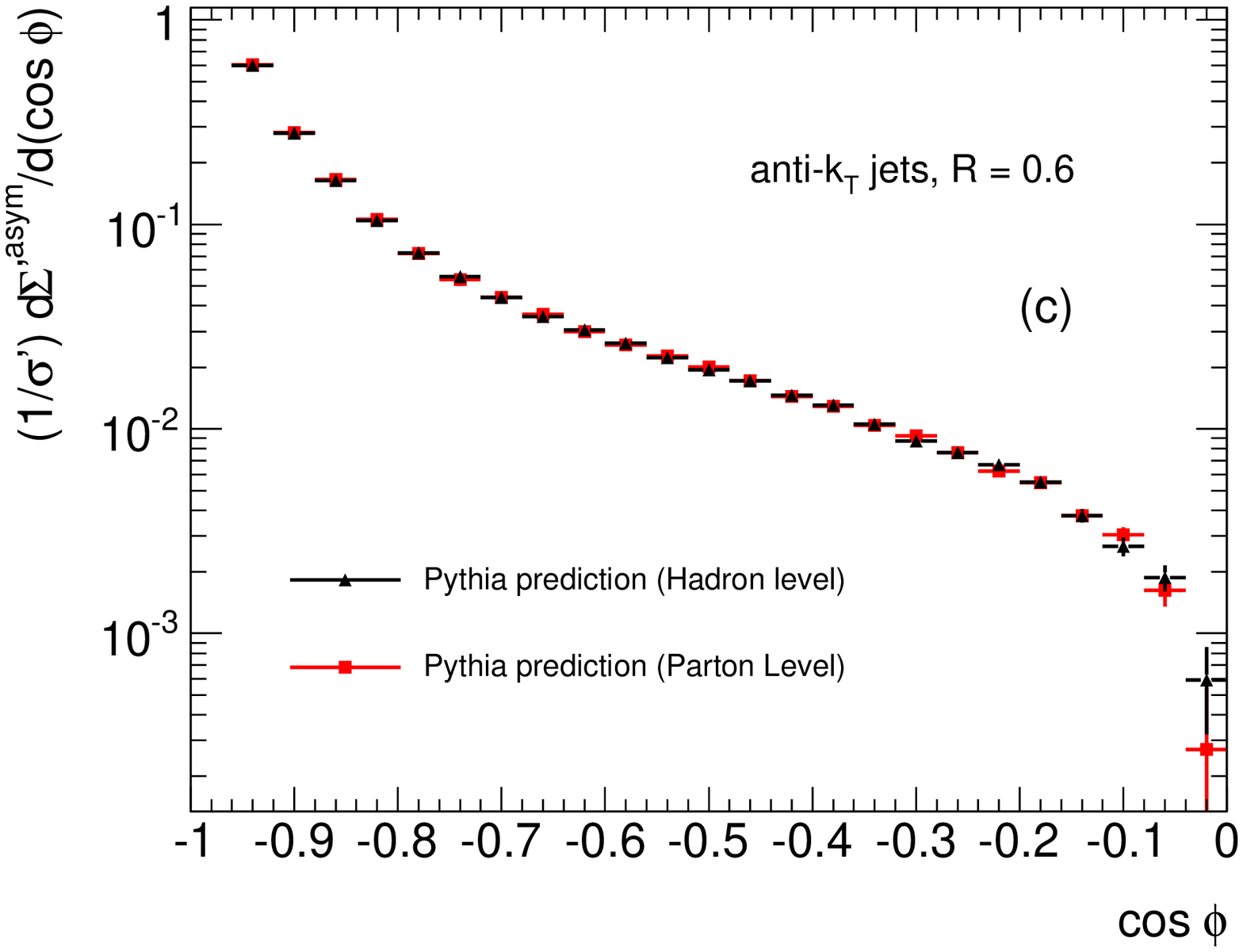}
\includegraphics[scale=0.44]{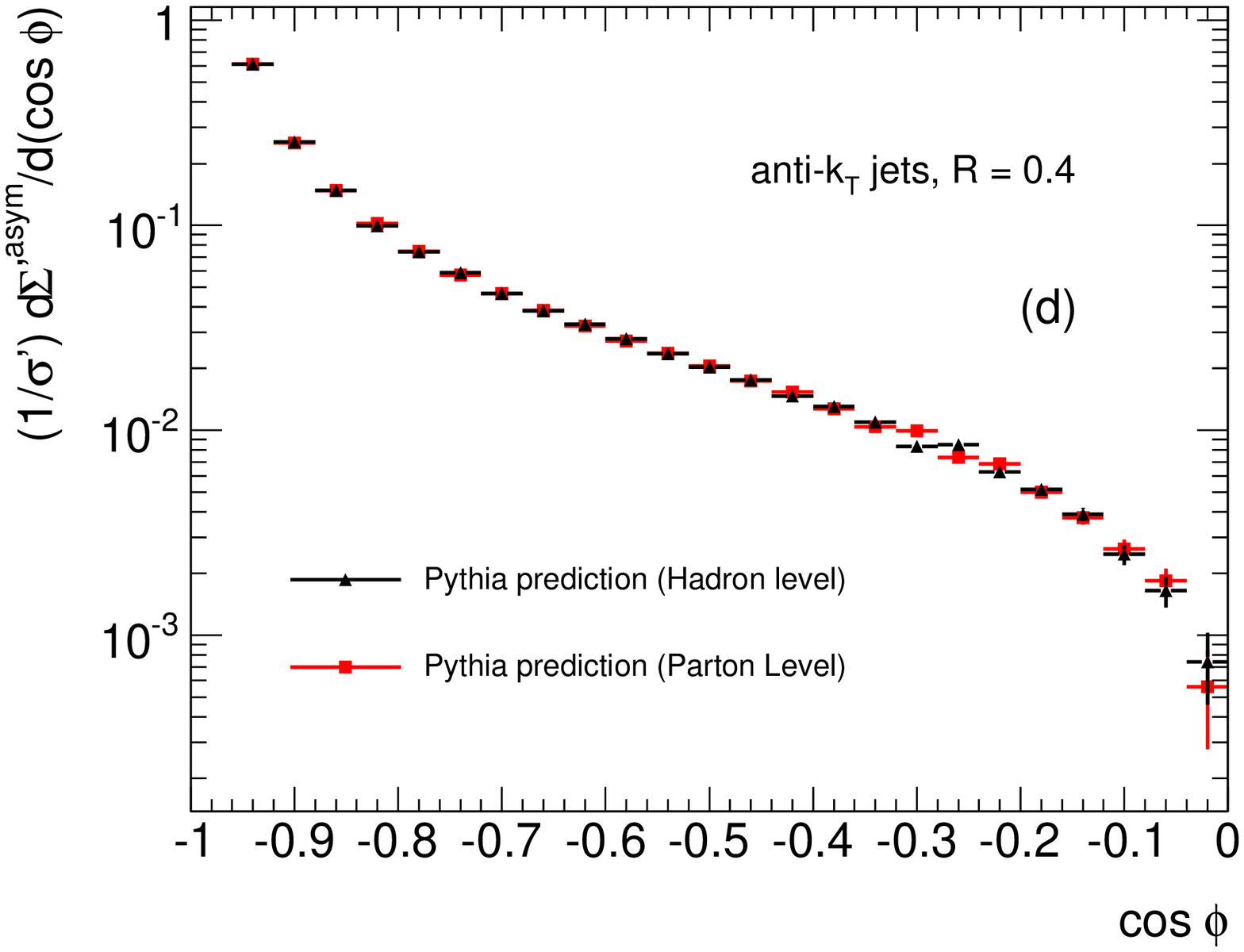} 
\caption{(color online) Differential distribution in $\cos\phi$ of the transverse EEC cross section [(a),(b)] and its asymmetry [(c),(d)] obtained with the  PYTHIA6 MC program~\cite{Sjostrand:2007gs}  at the hadron and parton level
at $\sqrt{s}=7$ TeV for the indicated values of the jet-size parameter $R$.  } \label{fig:hadron}
\end{center}
\end{figure}
%

\begin{figure}\begin{center}
\includegraphics[scale=0.44]{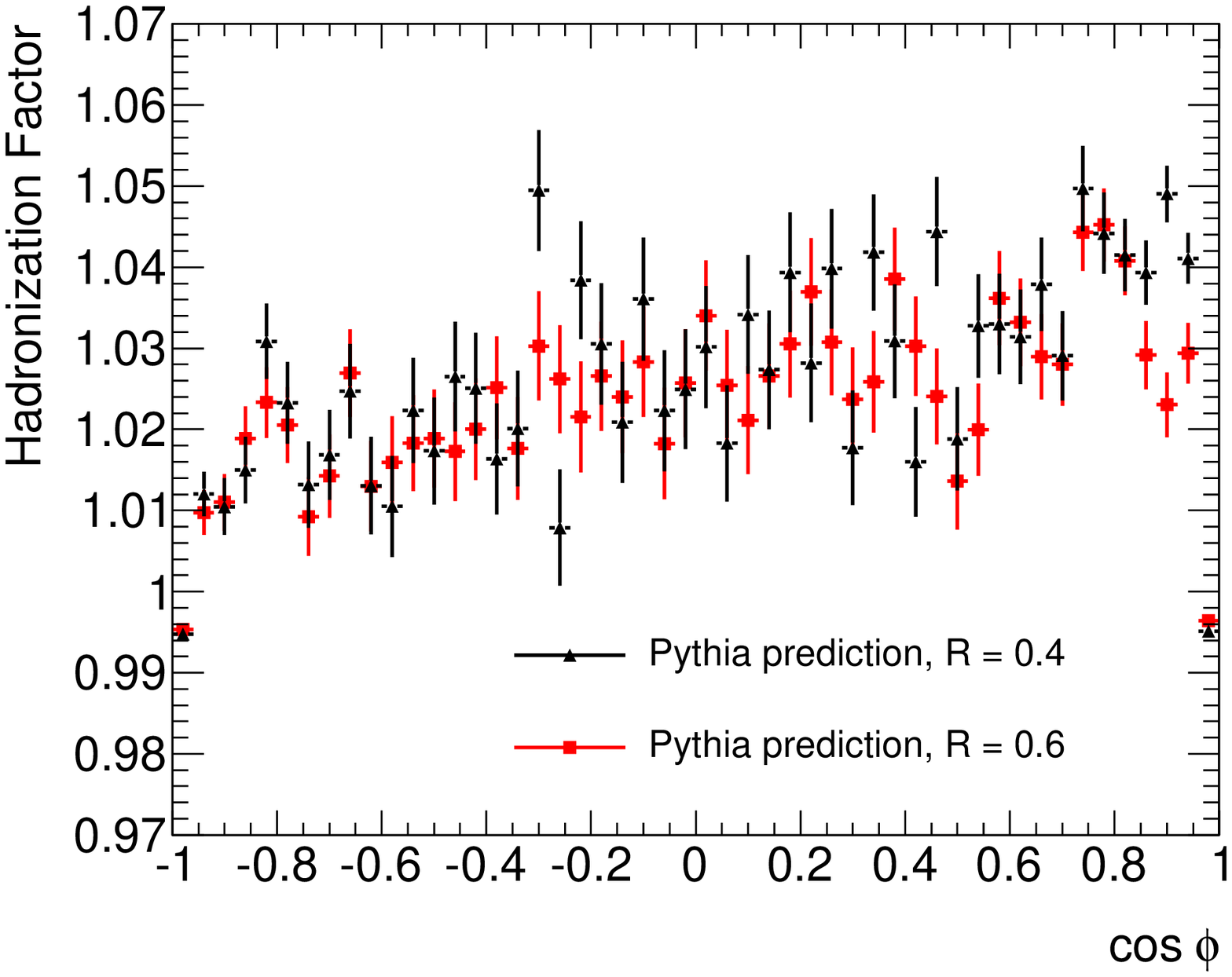}
\includegraphics[scale=0.44]{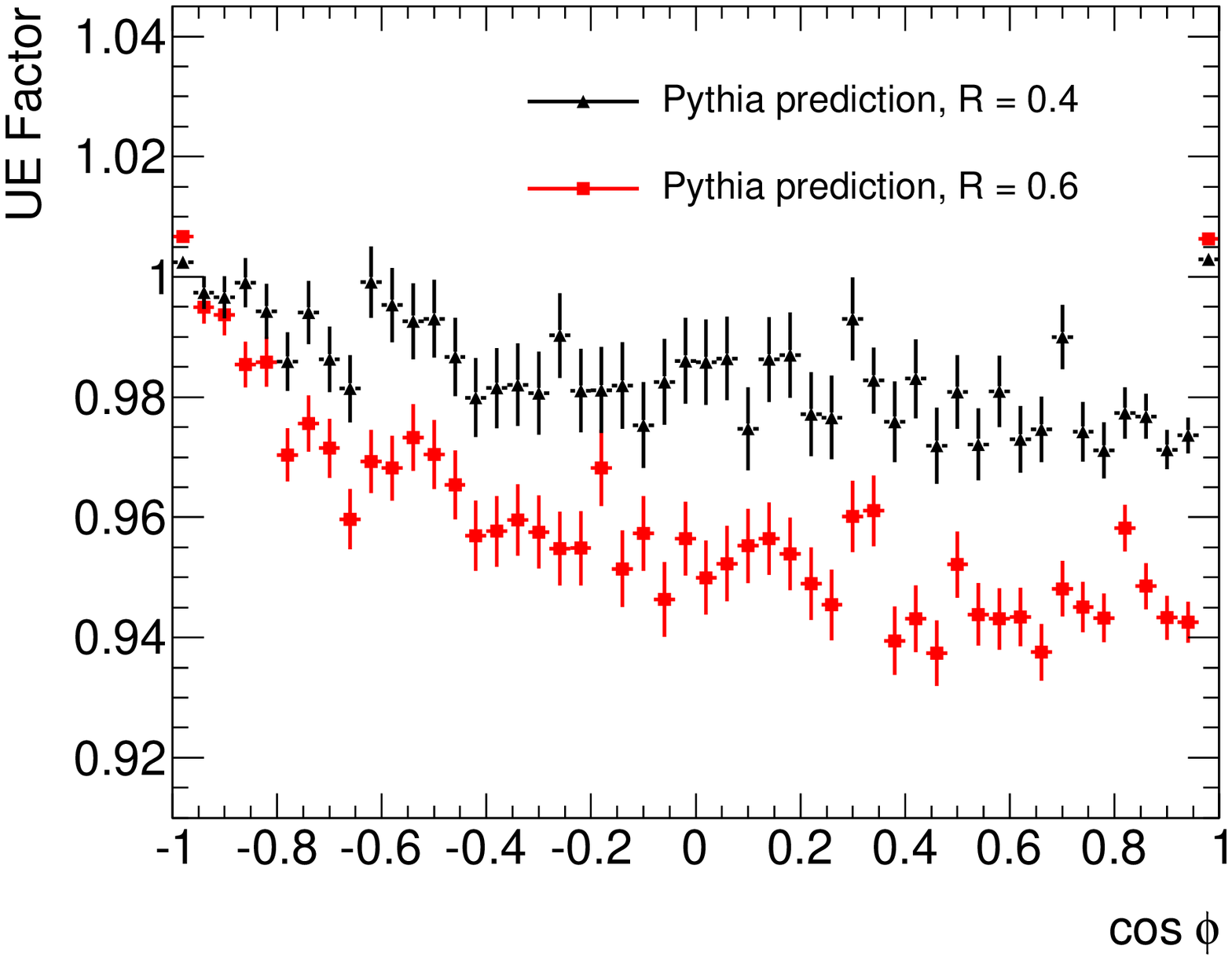}  
\caption{(color online) Normalized  distribution of the hadronization factor (left) and underlying  events effects (right) obtained with the  PYTHIA6 MC program~\cite{Sjostrand:2007gs} for the two indicated values of $R$.   } \label{fig:hadron_UE}
\end{center}
\end{figure}
%

The effects induced by the underlying event,  multiparton interactions  and hadronization effects have been
studied by us using the PYTHIA6 MC~\cite{Sjostrand:2007gs}.  
In Fig~.\ref{fig:UE}, we show a comparison of the transverse EEC and its asymmetry for R=0.6 and R=0.4
 with and without the underlying event effects (UE). In Fig~.\ref{fig:hadron},  the results of the transverse EEC
 and its asymmetry at the hadron and parton level are presente for R=0.6 and R=0.4. 
 To better display this,  
we show in Fig.~\ref{fig:hadron_UE} the  normalized distribution of the hadronization factor (left) and the underlying 
events factor (right), from which it is easy to see that both the hadronization and  the UE effects are small.
Typically, the effect of hadronization on the transverse EEC is $\le 5\%$ and from the underlying event $\le 6\%$
for the jet-size parameter $R=0.6$. The corresponding numbers are $\le 5\%$ and $\le  2\%$ for $R=0.4$.
The parameter specifying the jet-size in the anti-$k_T$ algorithm is chosen as $R=0.4$ in the rest of this paper,
 as this choice makes the transverse EEC distribution less sensitive to the underlying minimum-bias events.  Moreover,
 a smaller value of $R$ induces smaller distortions on the EEC distribution for the smaller values
of the angle $\phi$. 

 An important issue  is the effect of the
parton showers in the transverse EEC and the AEEC distributions. They are crucially important in the
 $\phi \to 0^\circ$ and $\phi \to 180^\circ$ angular regions, but their effect is expected to be small in the central
angular range on which we have concentrated. We have checked this (approximately)
by comparing the results in the LO accuracy with those from the parton shower-based MC generator
PYTHIA6~\cite{Sjostrand:2007gs}, which is accurate in the leading log approximation and also includes
some NLO terms. Matching the NLO computations with the
parton shower simulations in the complete
next-to-leading log (NLL) accuracy is the aim of several  approaches, such as
the  POWHEG method, pioneered and subsequently developed 
in~\cite{Nason:2004rx,Frixione:2007vw}, which would allow to quantitatively compute the 
end-point region in the transverse EEC cross section~\cite{matching}.
Likewise, resummed perturbative techniques have been developed in a number
of dedicated studies for some event shape variables in hadronic
collisions~\cite{Banfi:2004nk,Banfi:2004yd}, which would expand the domain of applicability of
the perturbative techniques to a wider angular region in $\phi$.

%
\begin{figure}\begin{center}
\includegraphics[scale=0.44]{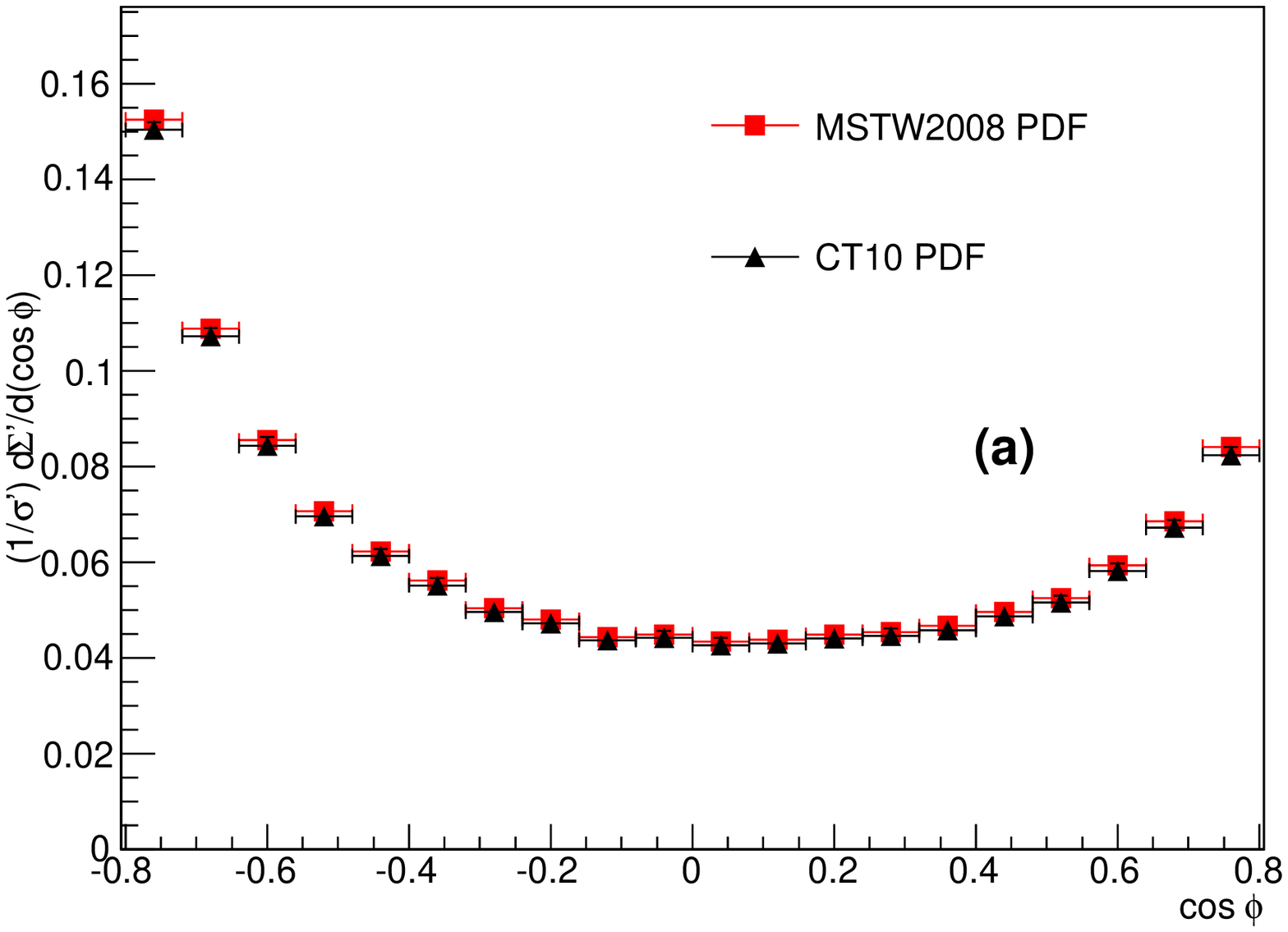}
\includegraphics[scale=0.44]{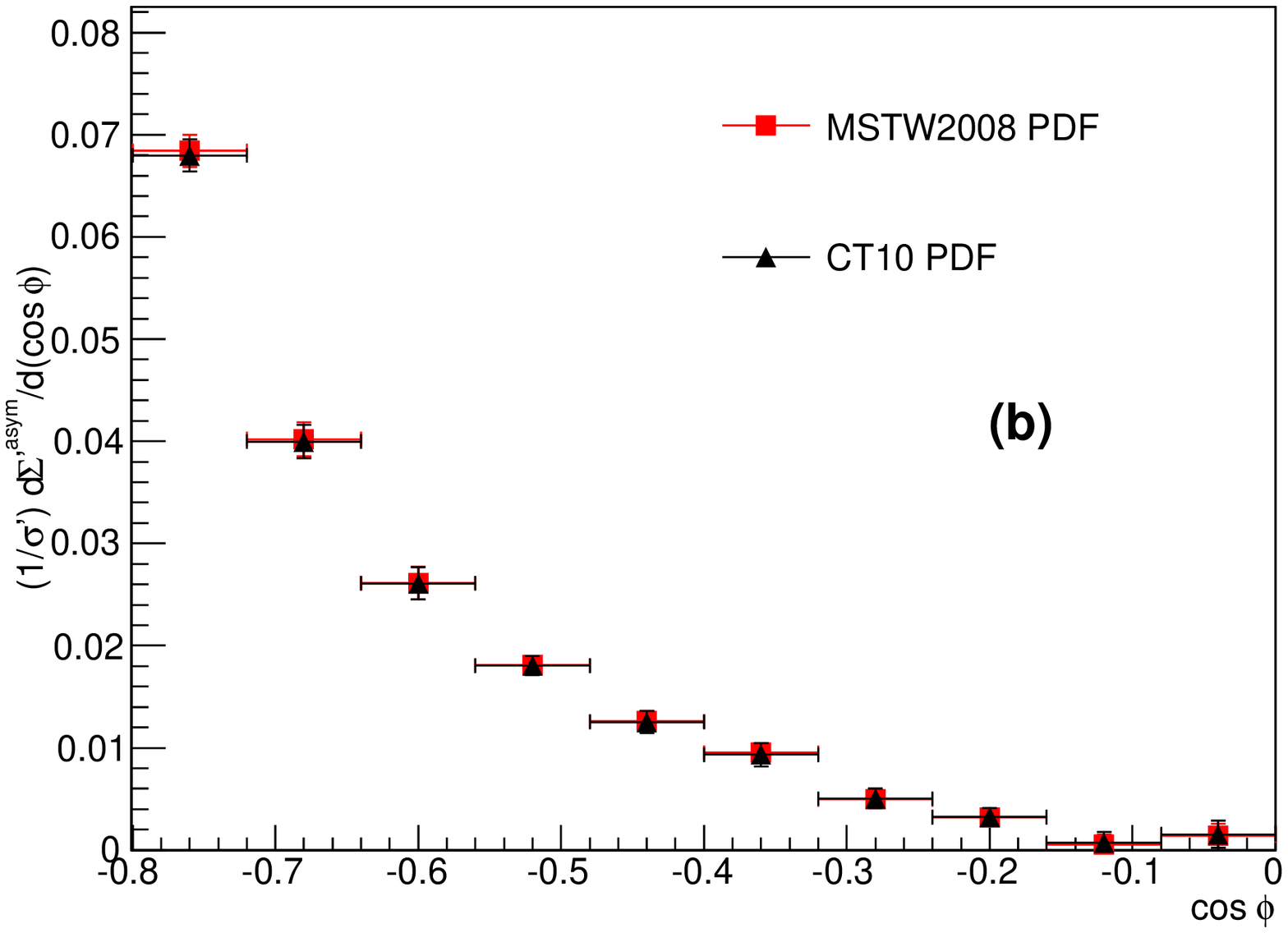} 
\caption{(color online) Dependence of the transverse EEC cross section (a) and its asymmetry  (b) on the PDFs at NLO  in $\alpha_s$. 
Red entries  correspond to the MSTW~\cite{Martin:2009iq} PDFs and the black ones are calculated using the
 CT10 PDF set~\cite{Lai:2010vv}.  The errors shown reflect the intrinsic parametric uncertainties in each  PDF set and the Monte Carlo integration uncertainties. 
 } \label{fig:pdf}
\end{center}
\end{figure}
In view of the preceding discussion,  we have restricted  $\cos\phi$ in the range $[-0.8, 0.8]$
which is sliced  into 20 bins for the presentation of our numerical results. 
We first show the dependence of the transverse EEC calculated in the NLO
accuracy on the PDFs in Fig.~\ref{fig:pdf} for the two widely used sets:
 MSTW~\cite{Martin:2009iq} and CT10~\cite{Lai:2010vv}, using their respective central (default) parameters.  
This figure shows that the PDF-related differences on the transverse EEC are negligible,
with the largest difference found in some bins amounting to $3\%$, (but typically they are $<1\%$).  We also remark that the intrinsic  uncertainties from the MSTW2008 PDFs, taking the first 10 eigenvectors of the PDF sets to evaluate the distributions,  are  found negligibly small in the transverse EEC (at most a few per mill), while in the case of CT10, these uncertainties are somewhat larger but still below $1\%$ in the EEC. 
The insensitivity of the transverse EEC cross section to the PDFs provides a
direct test of the underlying partonic hard processes.  In what follows, we will 
adopt the MSTW~\cite{Martin:2009iq} PDF set as it provides a correlated range of $\alpha_s(M_Z)$
and the structure functions for the current range of interest for
 $\alpha_s(M_Z)$: $0.11<\alpha_s(M_Z)<0.13$.

%
\begin{figure}\begin{center}
\includegraphics[scale=0.43]{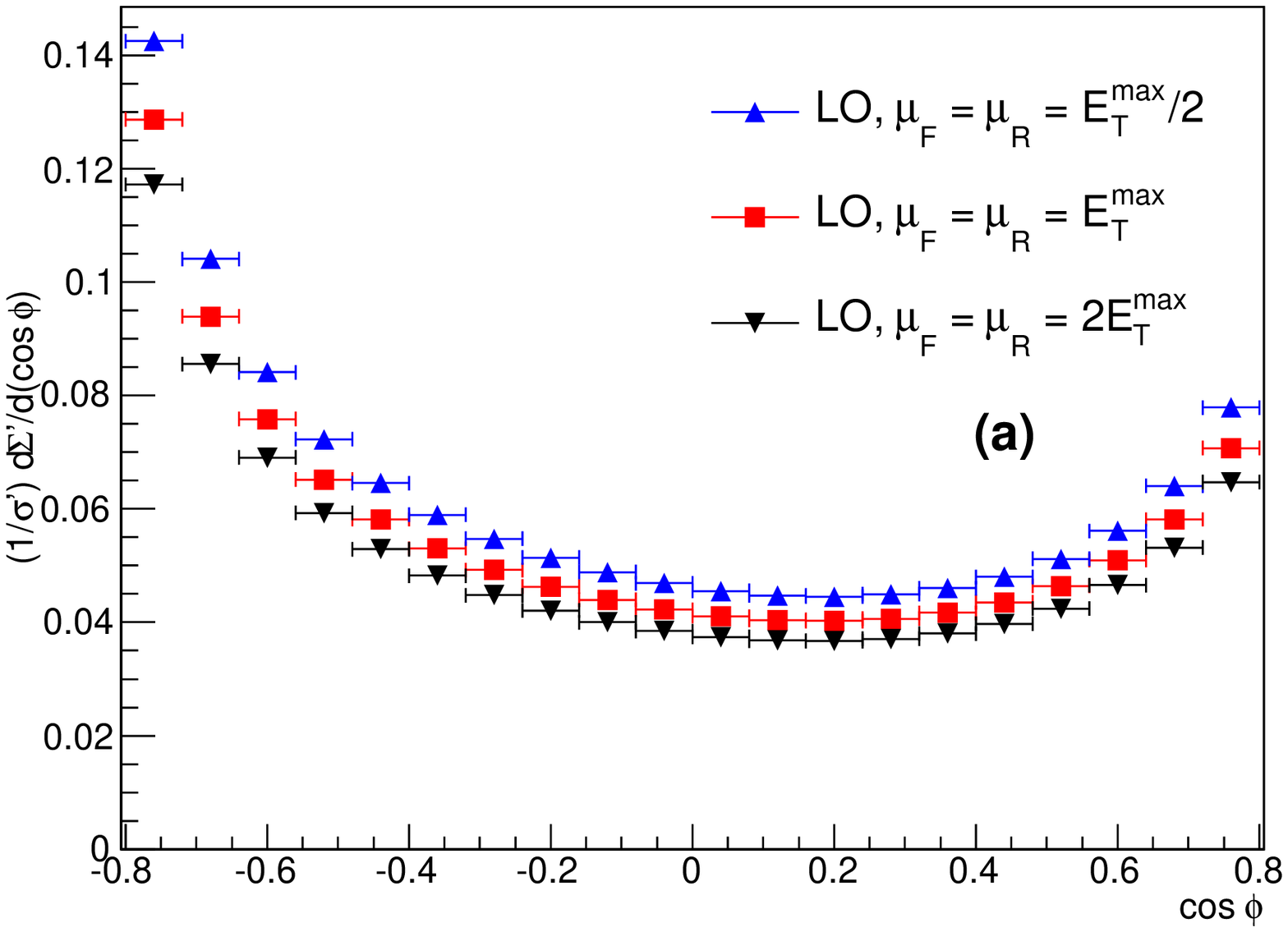}
\includegraphics[scale=0.43]{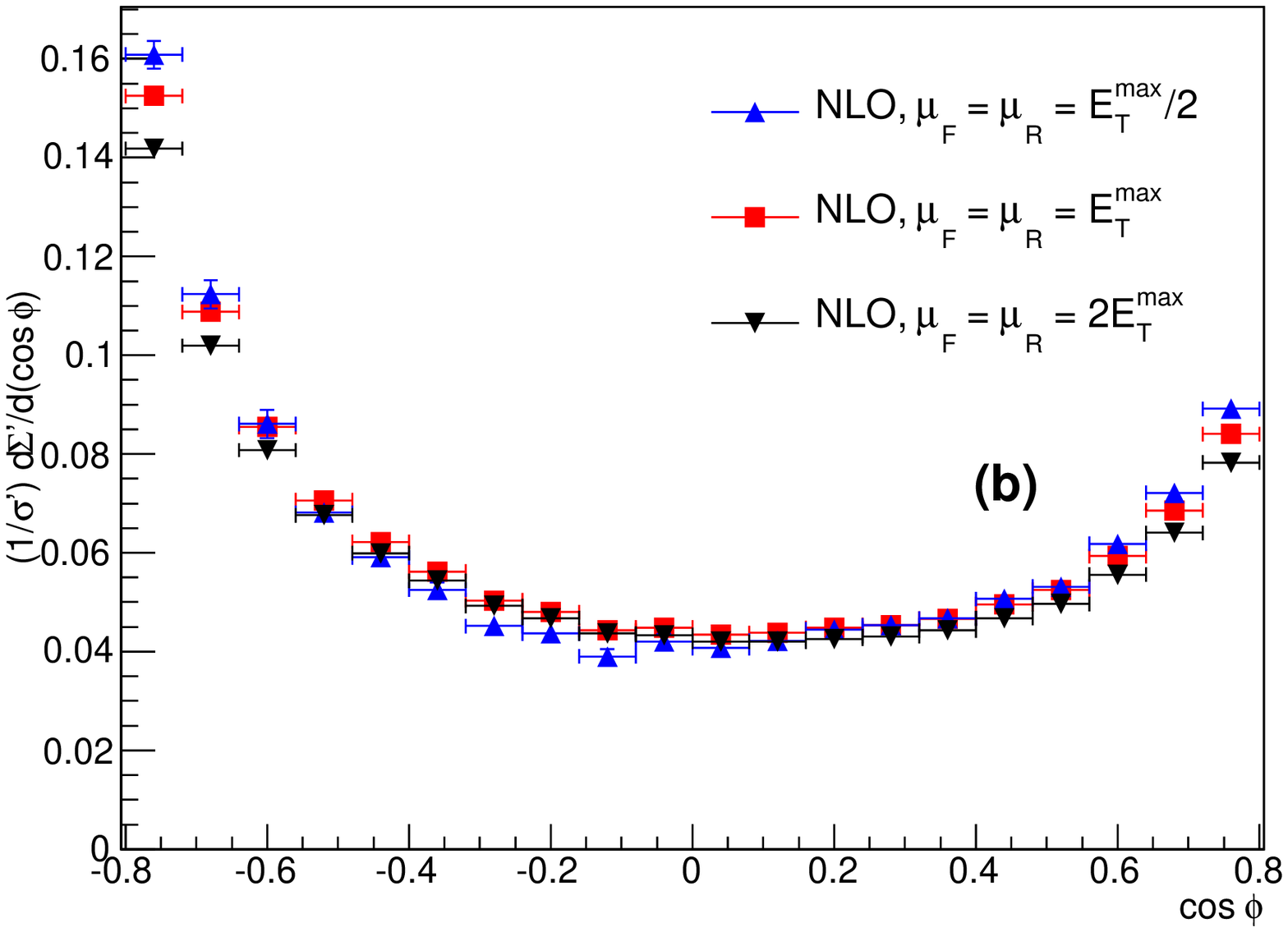}
\includegraphics[scale=0.43]{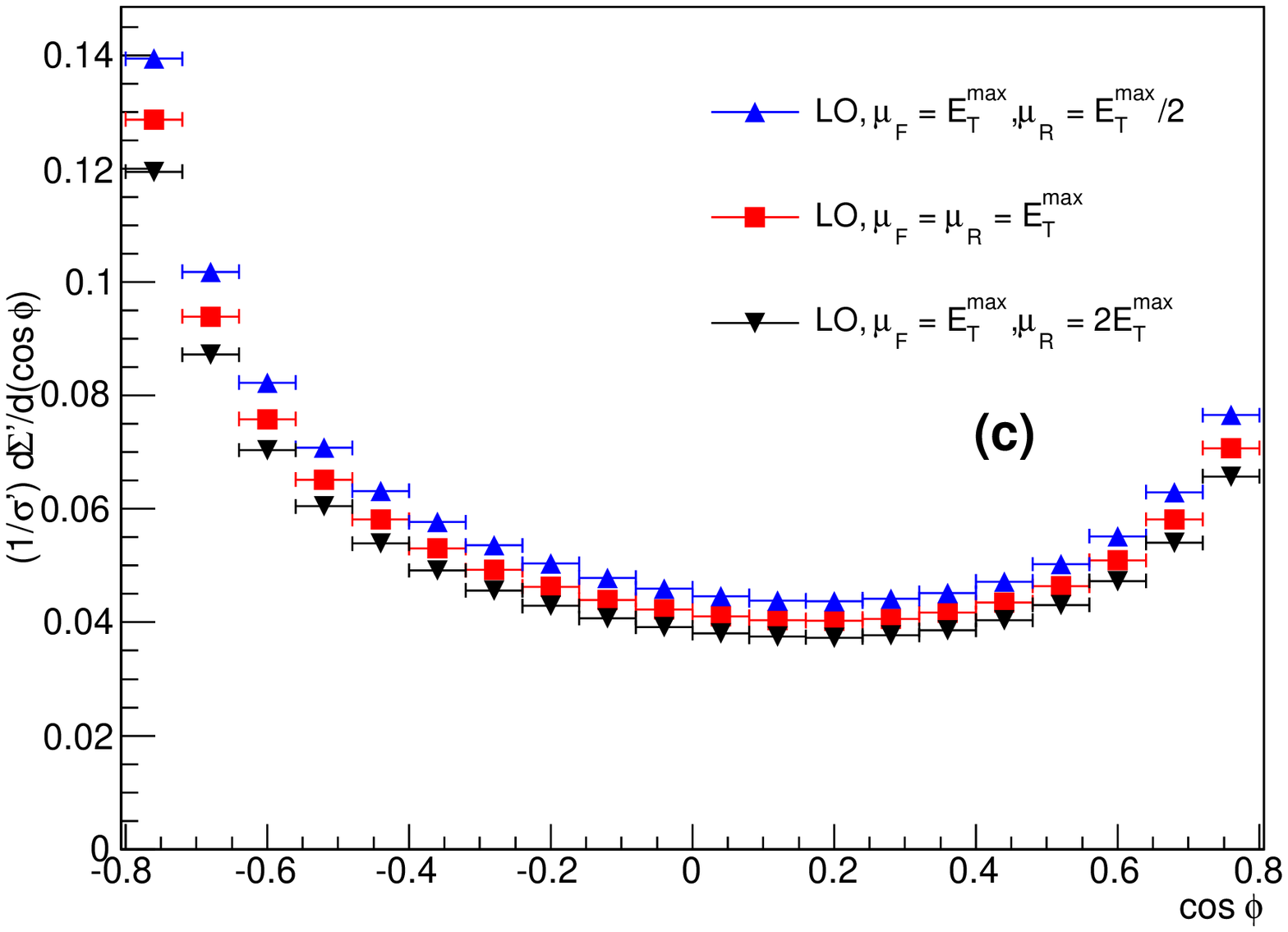}
\includegraphics[scale=0.43]{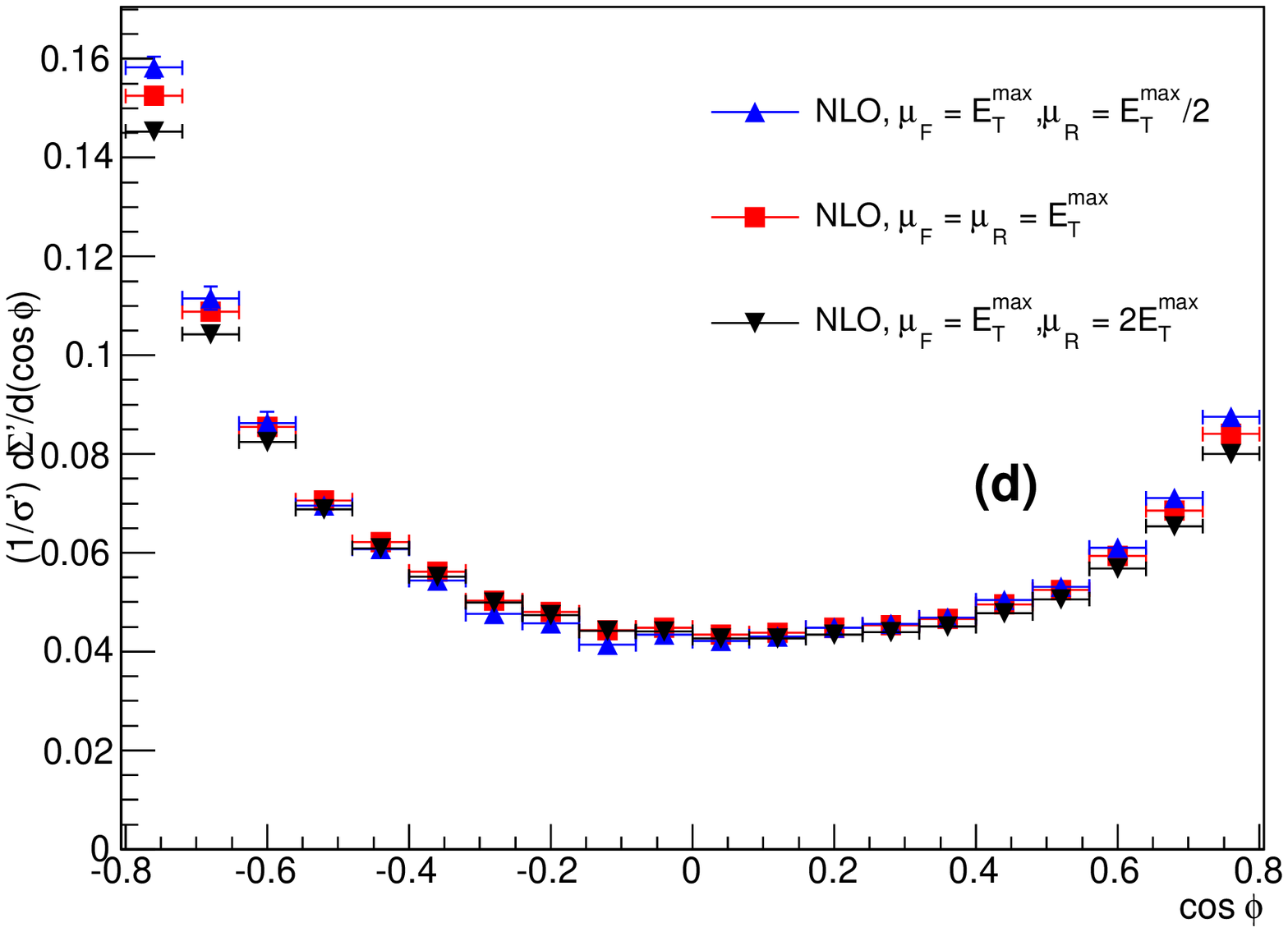}
\includegraphics[scale=0.43]{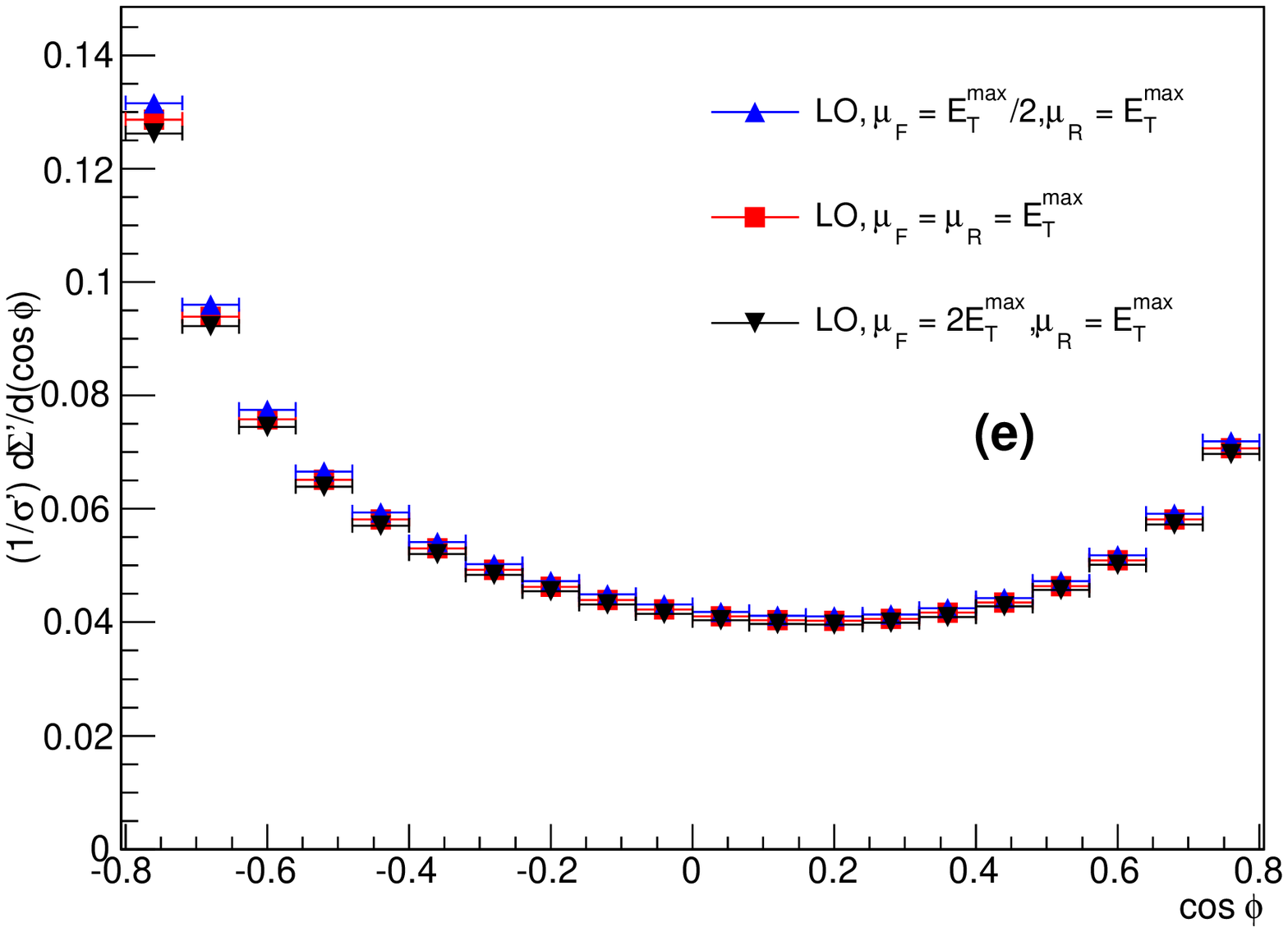}
\includegraphics[scale=0.43]{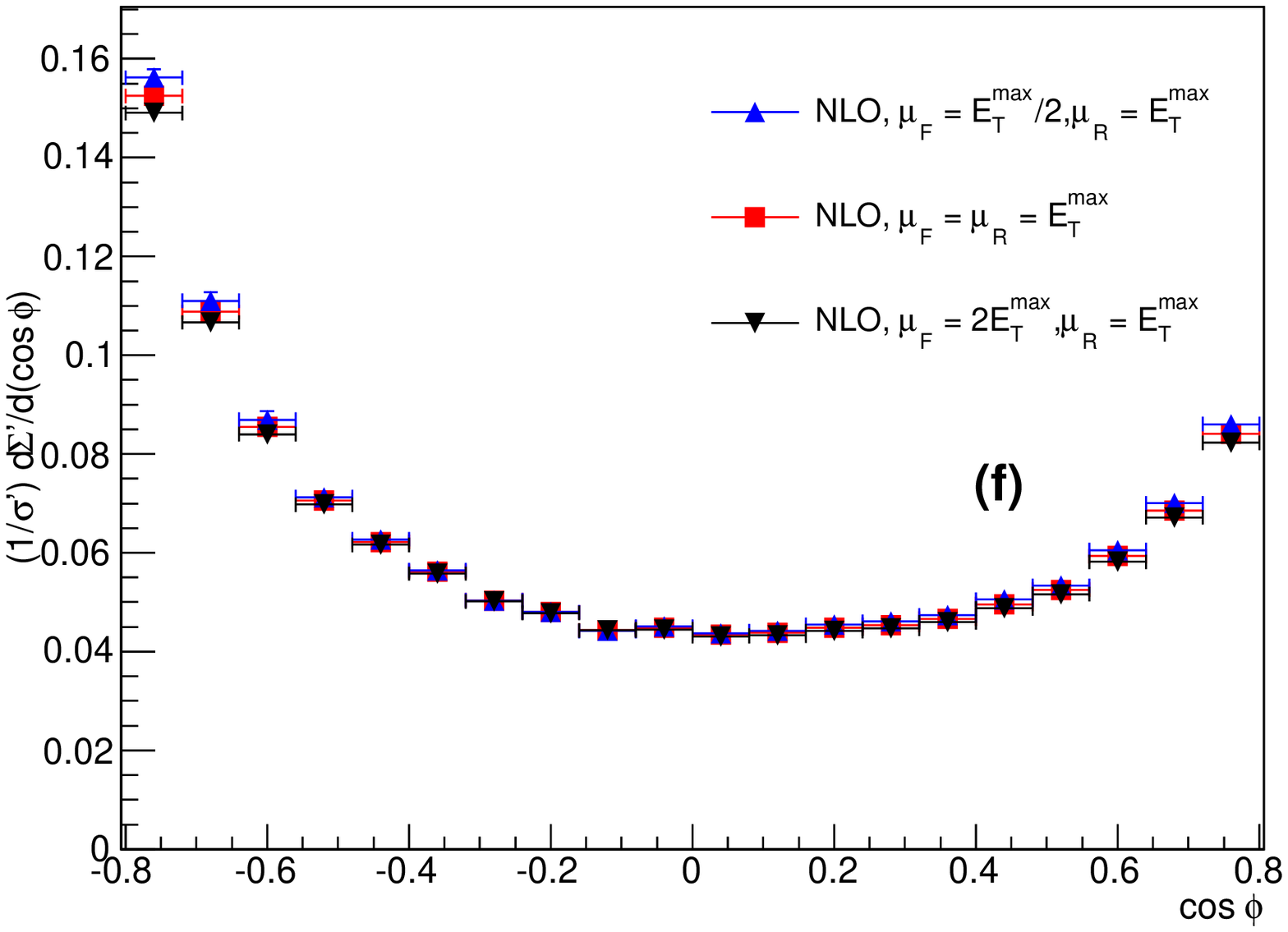}
\caption{ (color online) Dependence of the transverse EEC on the scales $\mu_F$, and $\mu_R$ in  LO (a,c,e) and in the NLO (b,d,f) in $\alpha_s$ for the
indicated values of the scales. Figs.(a) and (b) are obtained by setting $\mu_F=\mu_R$ and varying it $\mu_F=\mu_R= [0.5, 2]\times E_{T}^{\rm max}$; (c) and (d) are obtained  by fixing $\mu_F=E_{T}^{\rm max}$ and varying $\mu_R$, whereas  (e) and (f) are derived varying  $\mu_F$ with fixed $\mu_R=E_{T}^{\rm max}$. } \label{fig:scale-EEC}
\end{center}
\end{figure}

\begin{figure}\begin{center}
\includegraphics[scale=0.43]{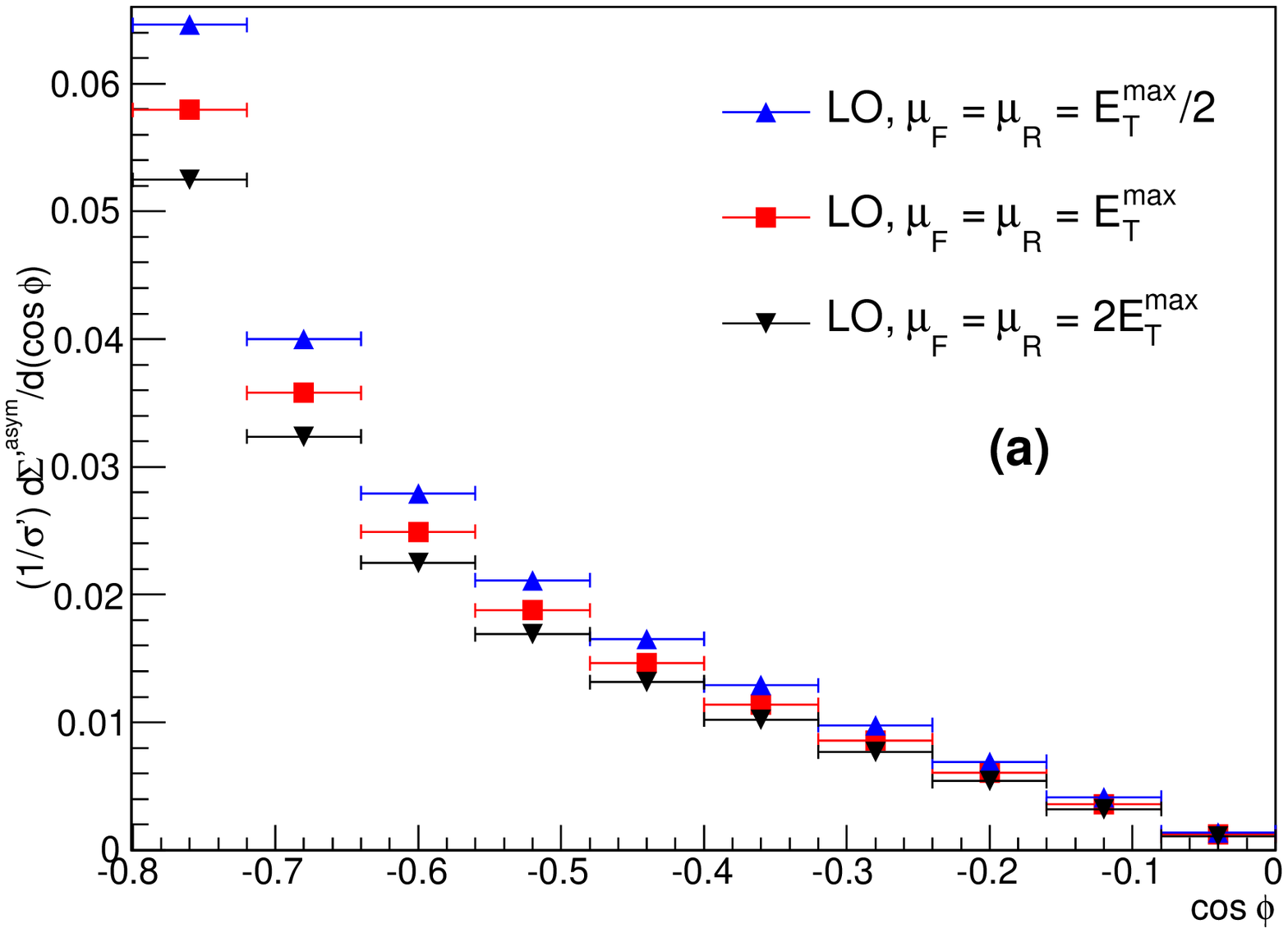}
\includegraphics[scale=0.43]{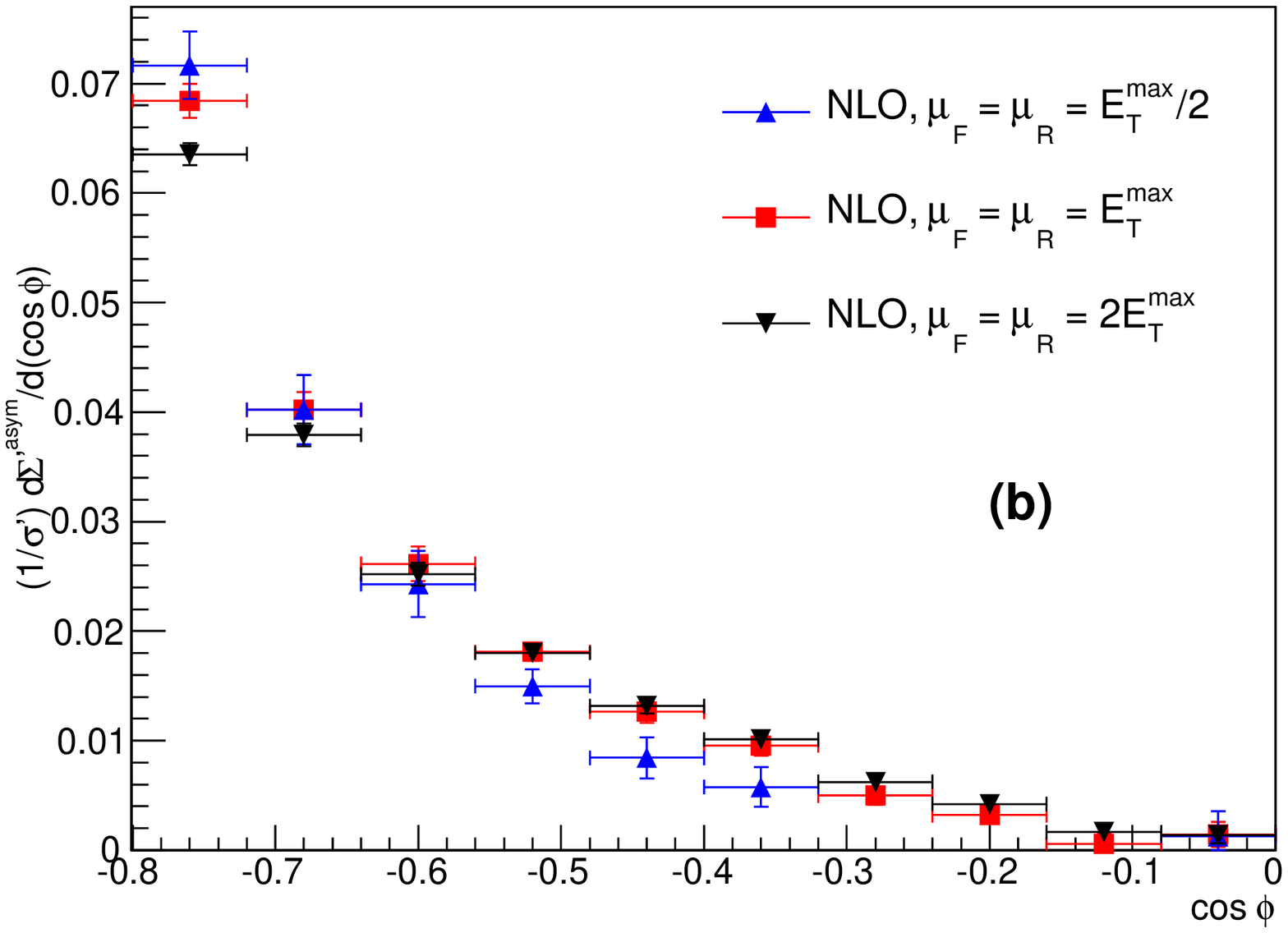}
\includegraphics[scale=0.43]{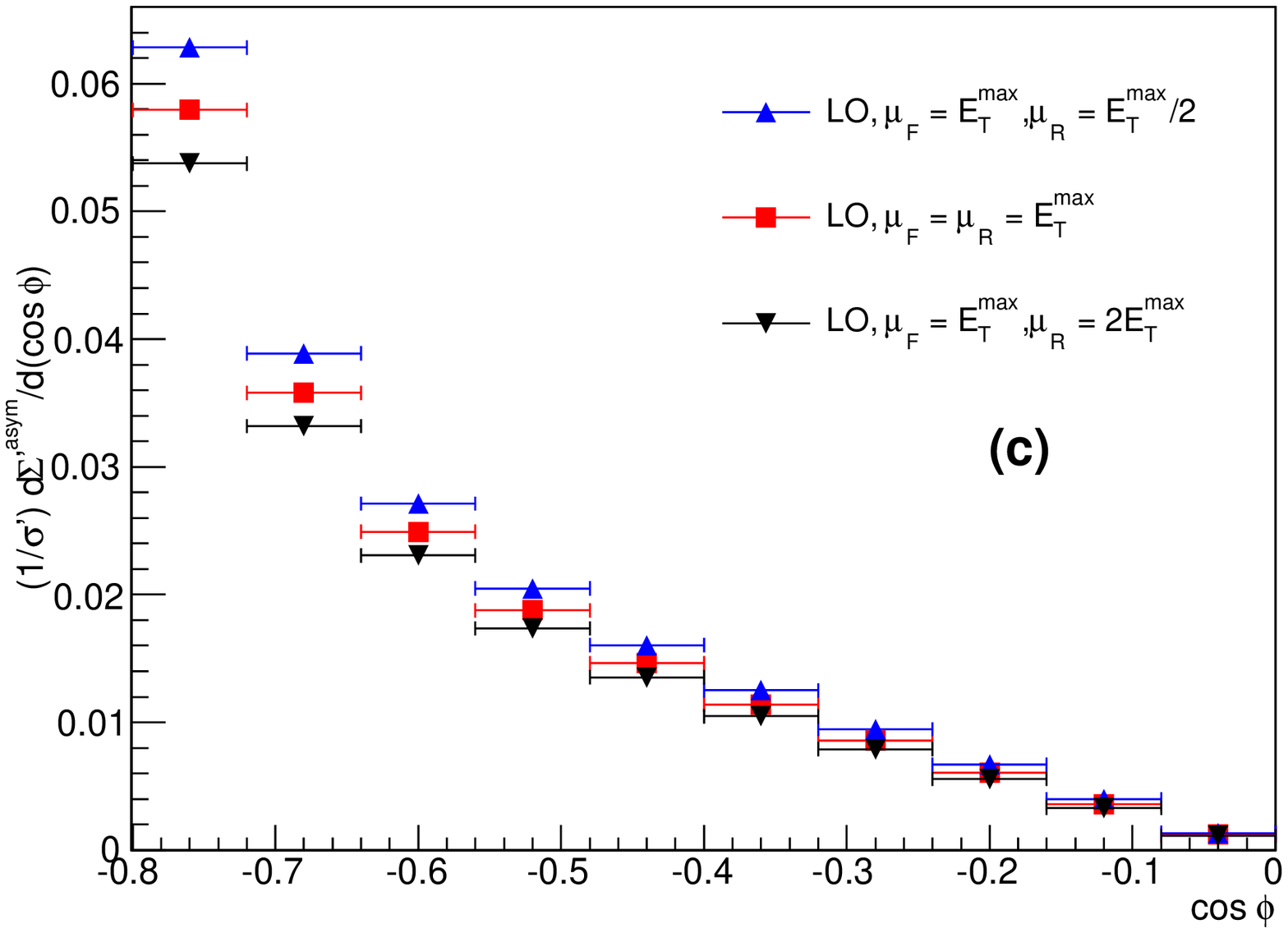}
\includegraphics[scale=0.43]{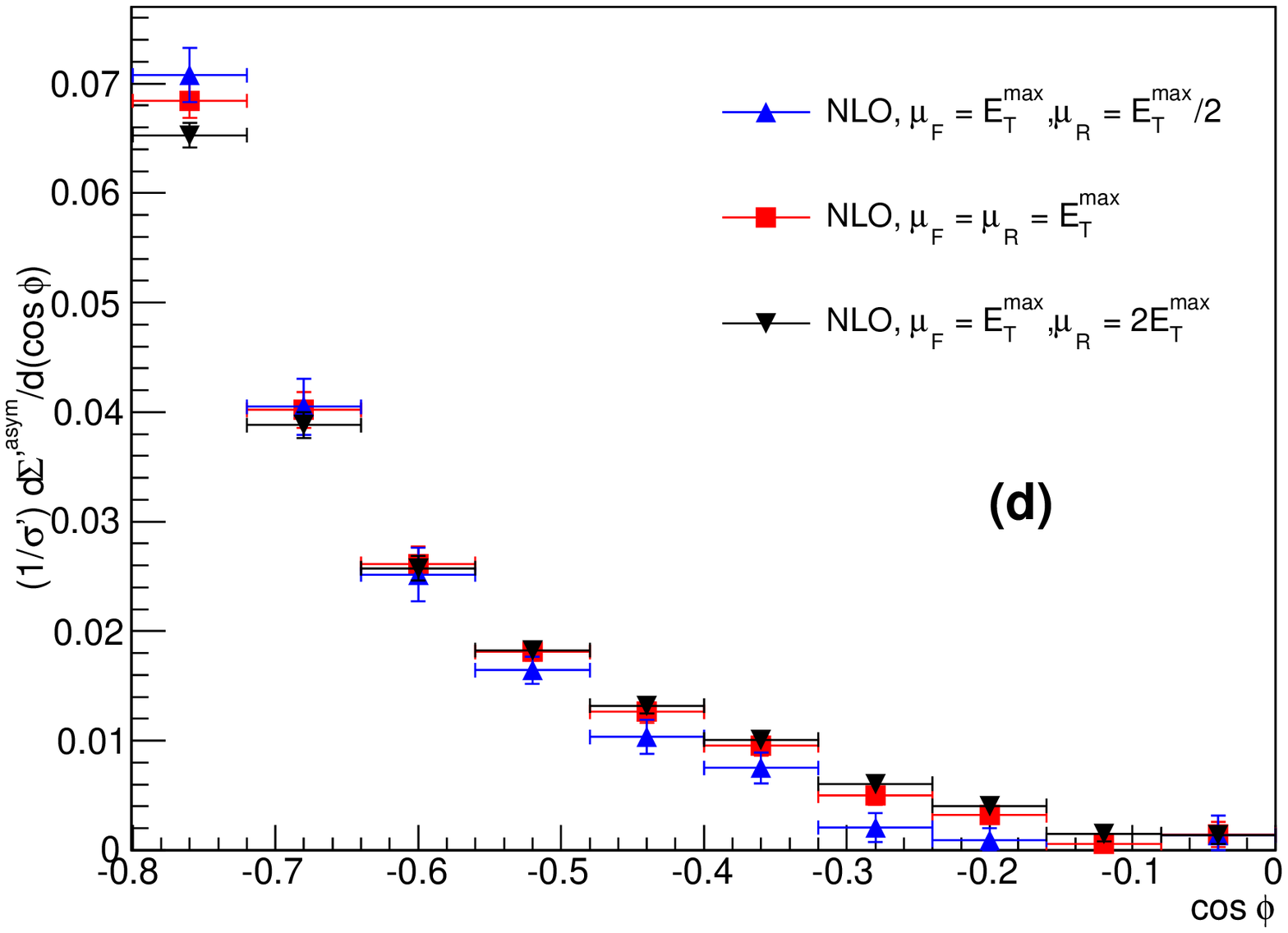}
\includegraphics[scale=0.43]{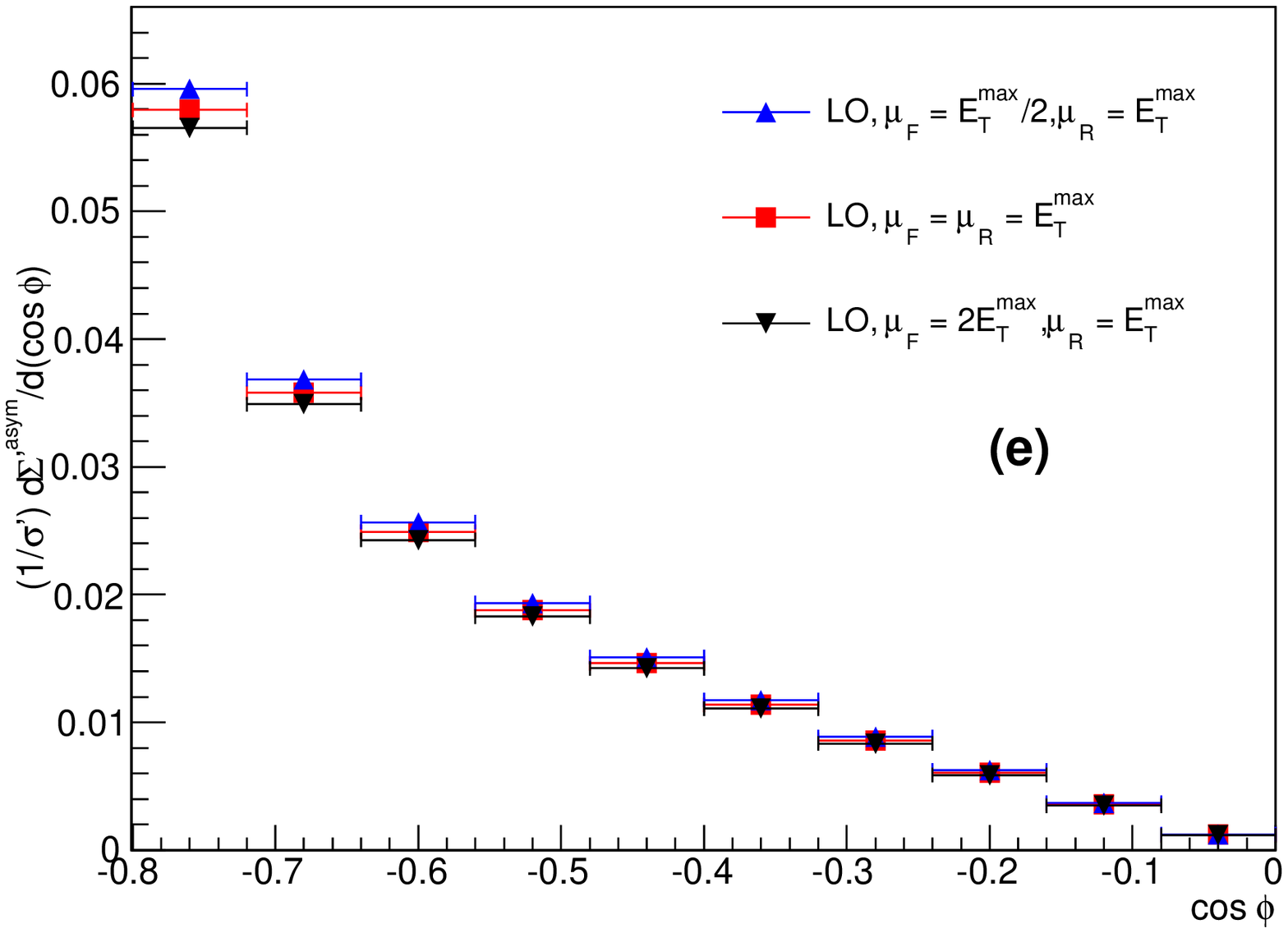}
\includegraphics[scale=0.43]{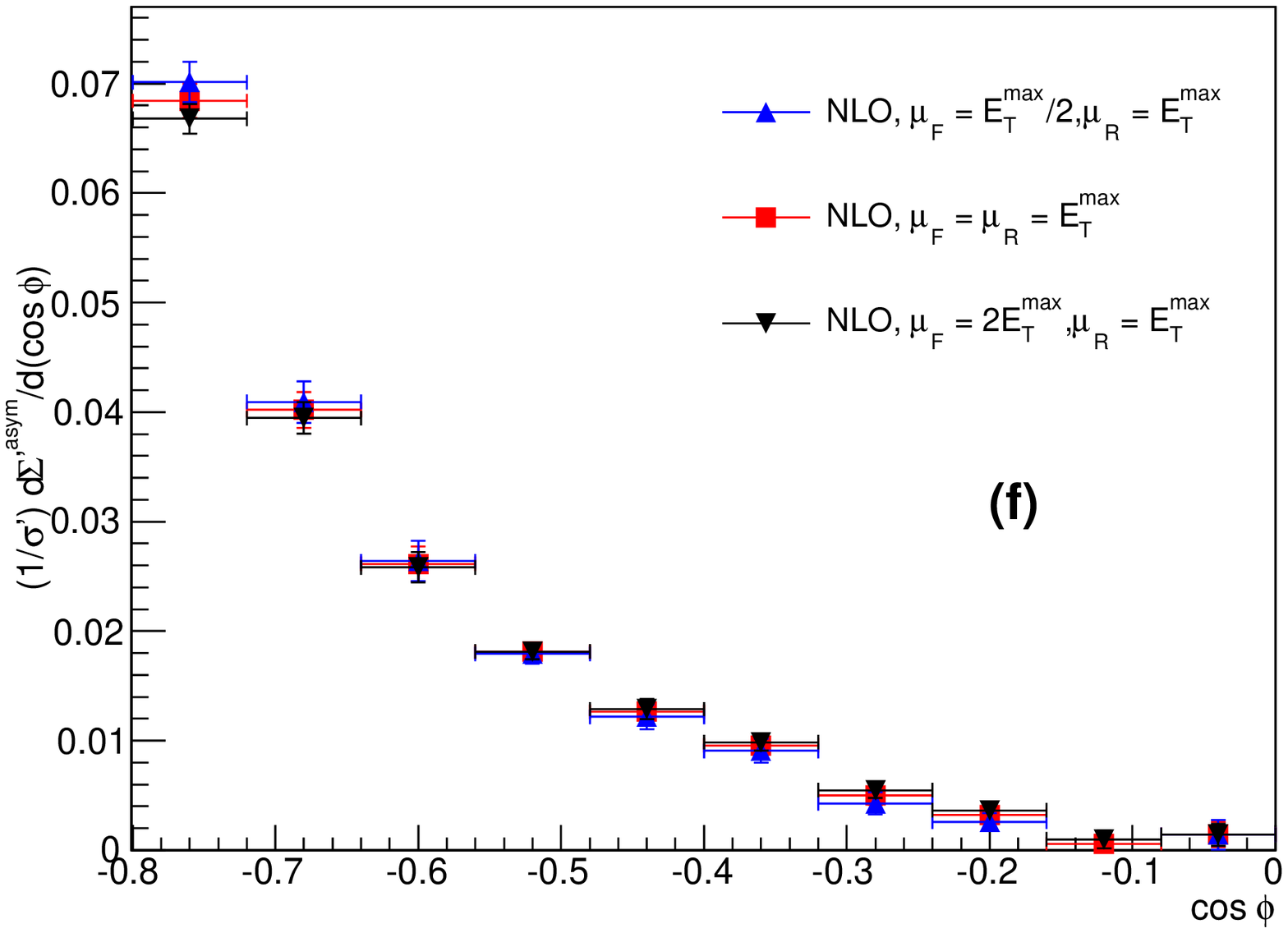}
\caption{(color online) Same as Fig.~\ref{fig:scale-EEC} but for the asymmetric transverse EEC.  } \label{fig:scale-AEEC}
\end{center}
\end{figure}

We next explore the dependences of the transverse EEC cross section  and its asymmetry
 on the factorization and the renormalization scales in the range
 $(\mu_F,\mu_R)= [0.5, 2]\times  E_{T}^{\rm max}$ and display them
 in Fig.~\ref{fig:scale-EEC} for the transverse EEC and Fig.~\ref{fig:scale-AEEC} for the asymmetric
 transverse EEC. 
Effects of the variations in the scales $\mu_F$ and $\mu_R$ on the transverse EEC cross section in the LO  are shown in Figs.~\ref{fig:scale-EEC} (a), \ref{fig:scale-EEC}(c) and  \ref{fig:scale-EEC} (e), which are obtained by setting the scales $\mu_F=\mu_R$, fixing $\mu_F= E_{T}^{\rm max}$ and varying $\mu_R$, and fixing $\mu_R= E_{T}^{\rm max}$ and varying $\mu_F$, respectively. The corresponding asymmetry of the transverse EEC cross sections are displayed in
 Figs.~\ref{fig:scale-AEEC} (a), \ref{fig:scale-AEEC} (c) and \ref{fig:scale-AEEC} (e). We note that the dominant scale dependence in the LO arises from the variation of the renormalization scale $\mu_R$. This is understandable as the LO matrix element have no $\mu_R$-compensating contribution. The results obtained in the NLO are shown in Figs.\ref{fig:scale-EEC} (b), \ref{fig:scale-EEC} (d) and \ref{fig:scale-EEC} (f) for the
 transverse EEC and in Figs. \ref{fig:scale-AEEC} (b), \ref{fig:scale-AEEC} (d) and \ref{fig:scale-AEEC} (f) for the asymmetry. One observes significantly less dependence  on the scales; in particular the marked $\mu_R$-dependence in the LO is now reduced. 
Typical scale-variance on the transverse EEC distribution in the NLO
is found to be  2\% - 3\%,  with the largest effects in some bins reaching $5\%$.
This scale-insensitivity in the NLO accuracy  is crucial to undertake
a quantitative determination of $\alpha_s$ from the collider jet data.

\begin{figure}\begin{center}
\includegraphics[scale=0.44]{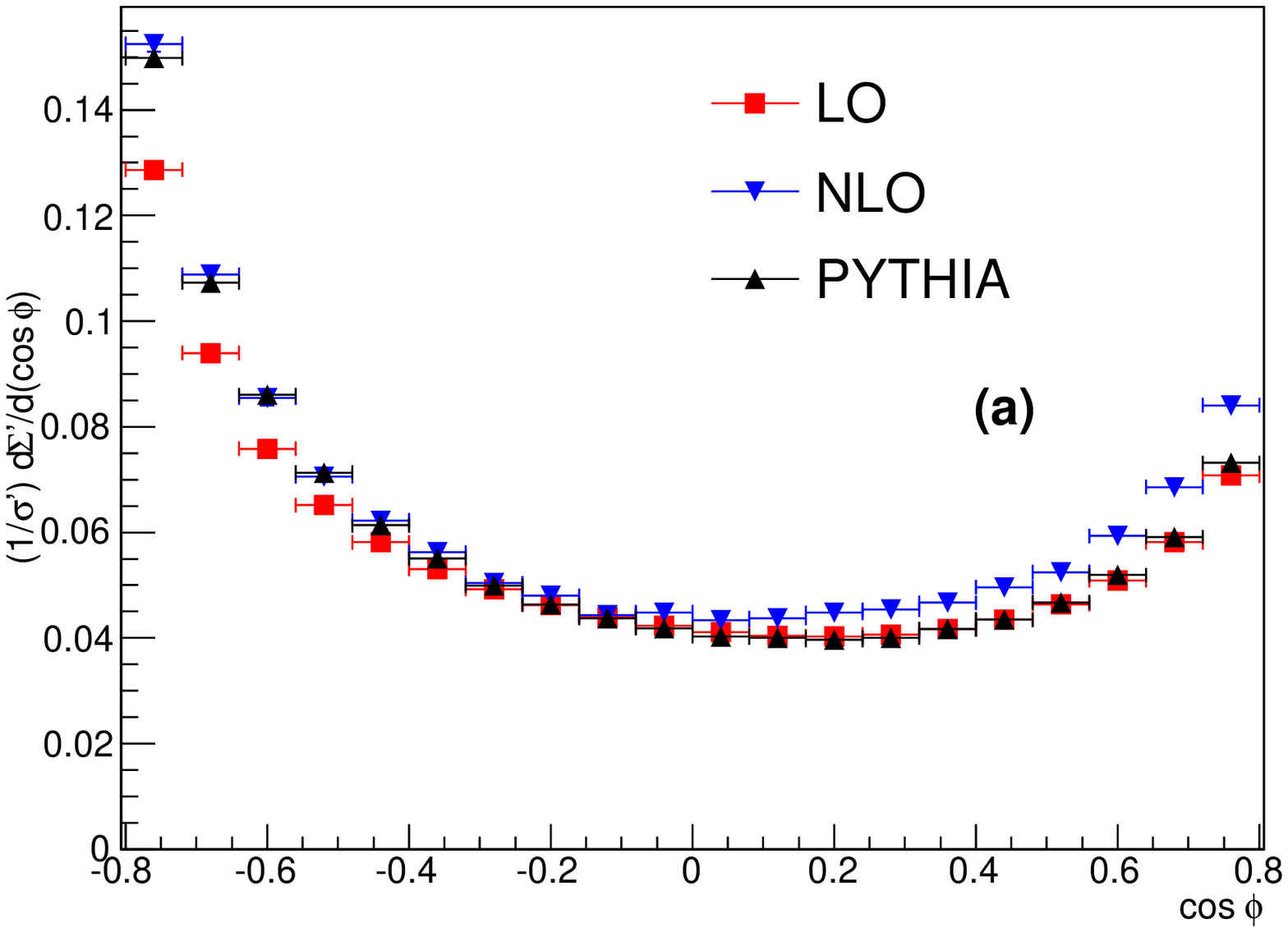}
\includegraphics[scale=0.44]{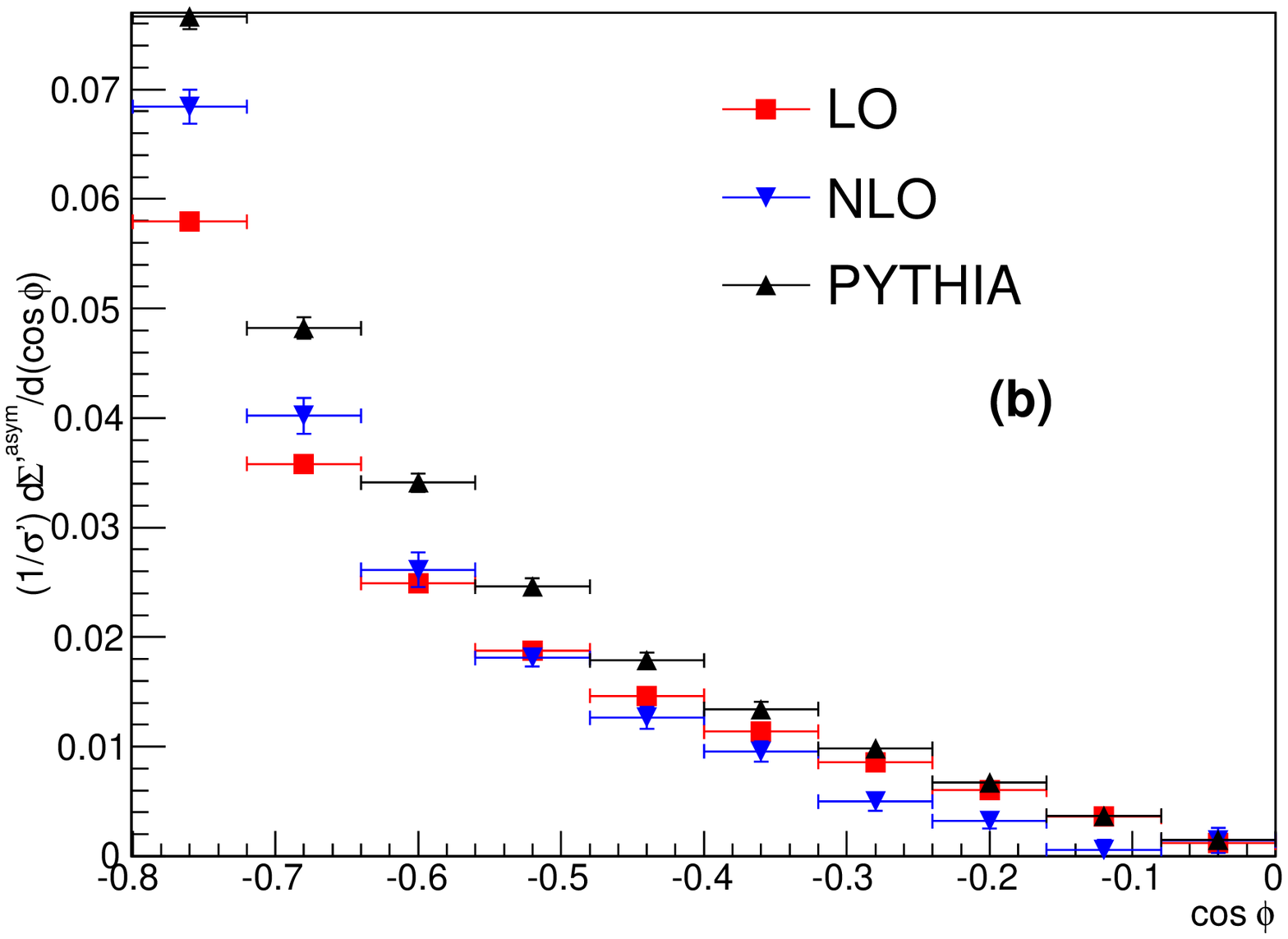}
\includegraphics[scale=0.44]{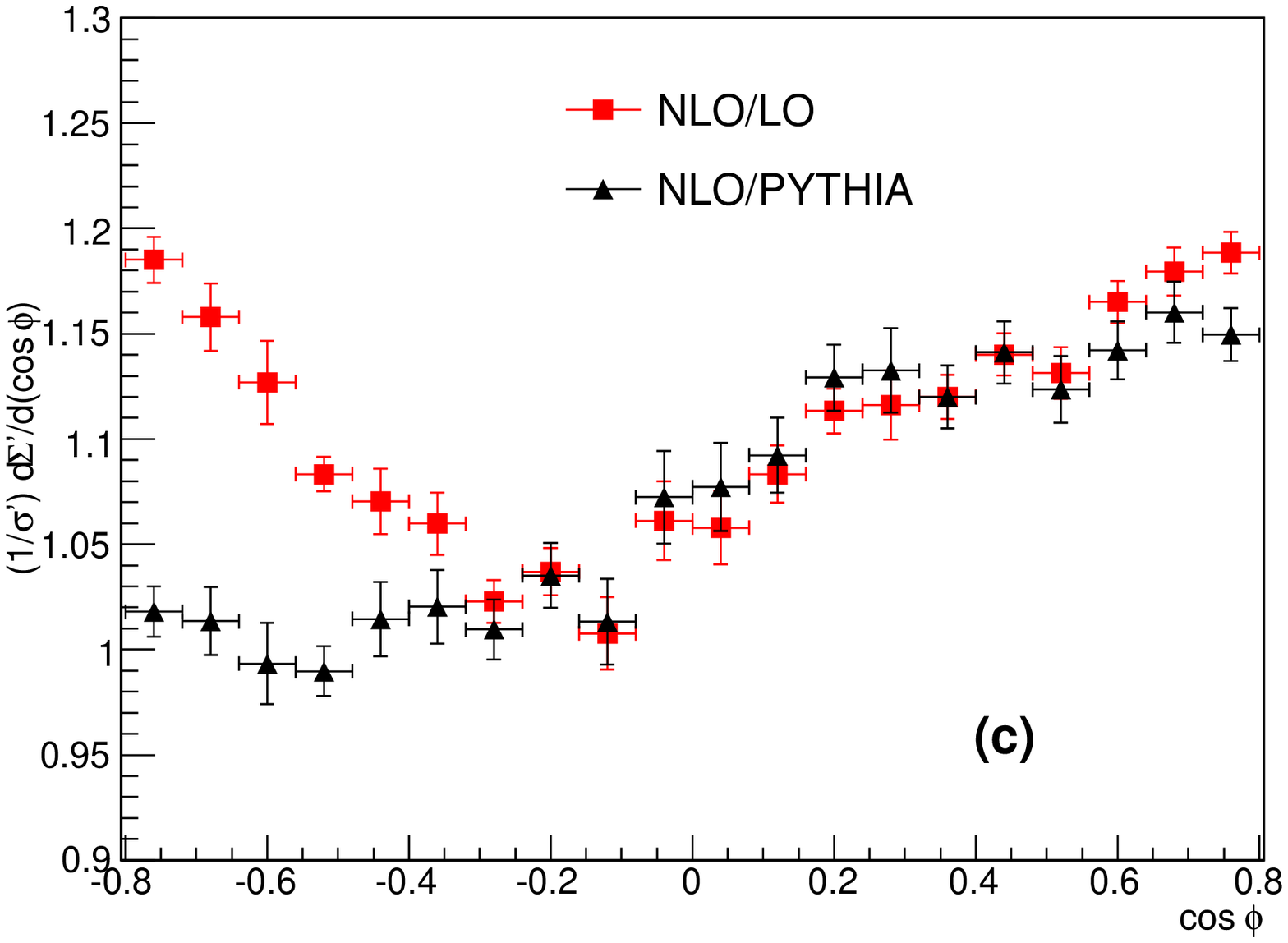}
\includegraphics[scale=0.44]{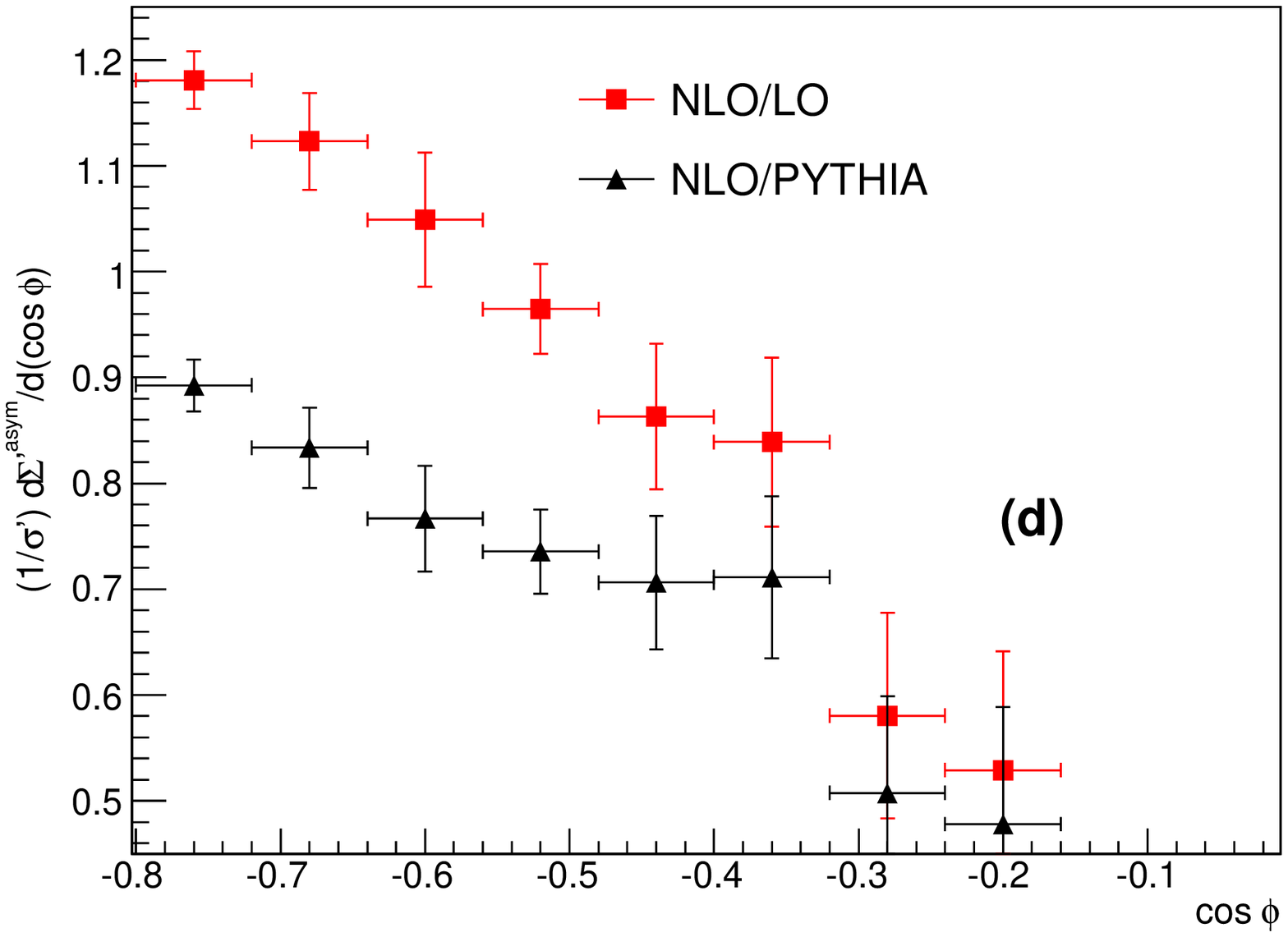}
\caption{(color online) Comparison of  the LO computation (red entries), NLO calculation (blue entries) and the
 PYTHIA (black entries) results  for the transverse EEC (a) and its asymmetry (b).
Fig. (c) displays the  function $K^{\rm EEC}(\phi)$ involving the ratios NLO/LO (red entries) as defined in Eq.~\eqref{eq:K-def} and a phenomenological function obtained by replacing the LO results by the PYTHIA MC results (black entries). Fig. (d) shows the corresponding   function $K^{\rm AEEC}(\phi)$ for the transverse EEC asymmetry defined in Eq.~\eqref{eq:K-AEEC-def} . The errors shown are obtained by adding  in quadrature all the uncertainties except the ones from scale variations, as described in text. 
 } \label{fig:comparison}
\end{center}
\end{figure}
%

Having shown that the uncertainties due to underlying events and the PDFs are negligible, and the scale dependence
 is much reduced in the NLO, we 
present our results for the transverse EEC in the LO  and the
NLO accuracy  in Fig.~\ref{fig:comparison}(a),
and the corresponding results for the transverse AEEC  in  Fig.~\ref{fig:comparison} (b).
We also compute these distributions from a MC-based model which has the LO matrix elements
and multiparton showers encoded. To be specific, we have used the PYTHIA8~\cite{Sjostrand:2007gs}
 MC program and have 
generated the transverse EEC and the AEEC distributions, which are also shown 
in  Fig.~\ref{fig:comparison}(a) and Fig.~\ref{fig:comparison}(b), respectively.
This comparison provides a practically convenient way to
correct the PYTHIA8 MC-based theoretical distributions, often used in the analysis of the
hadron collider data, due to the NLO effects.
 In  Fig.~\ref{fig:comparison} (c), we show the function $K^{\rm EEC}(\phi)$ 
defined in Eq.~(\ref{eq:K-def}) (denoted as NLO/LO in the figure) and another phenomenological  function
 in which the NLO
transverse EEC distribution is normalized to the one generated by the
 PYTHIA8~\cite{Sjostrand:2007gs} MC program (denoted as NLO/PYTHIA). The corresponding function $K^{\rm AEEC}(\phi)$, defined in Eq.~(\ref{eq:K-AEEC-def}), is shown in Fig.~\ref{fig:comparison} (d). Here also we show the corresponding phenomenological  function in which the transverse EEC obtained in NLO is normalized to the ones generated by the PYTHIA MC. 
 We remark that the effects of the NLO corrections
are discernible, both compared to the LO and PYTHIA8~\cite{Sjostrand:2007gs}, and they are significant in the
large-angle region (i.e., for $\cos \phi < 0$).
To summarize the NLO effects in the EEC distribution, they reduce the scale-dependence,  in particular
on $\mu_R$, and distort the shape of both the EEC and AEEC distributions, providing a non-trivial test of
the NLO effects.

\begin{figure}\begin{center}
\includegraphics[scale=0.44]{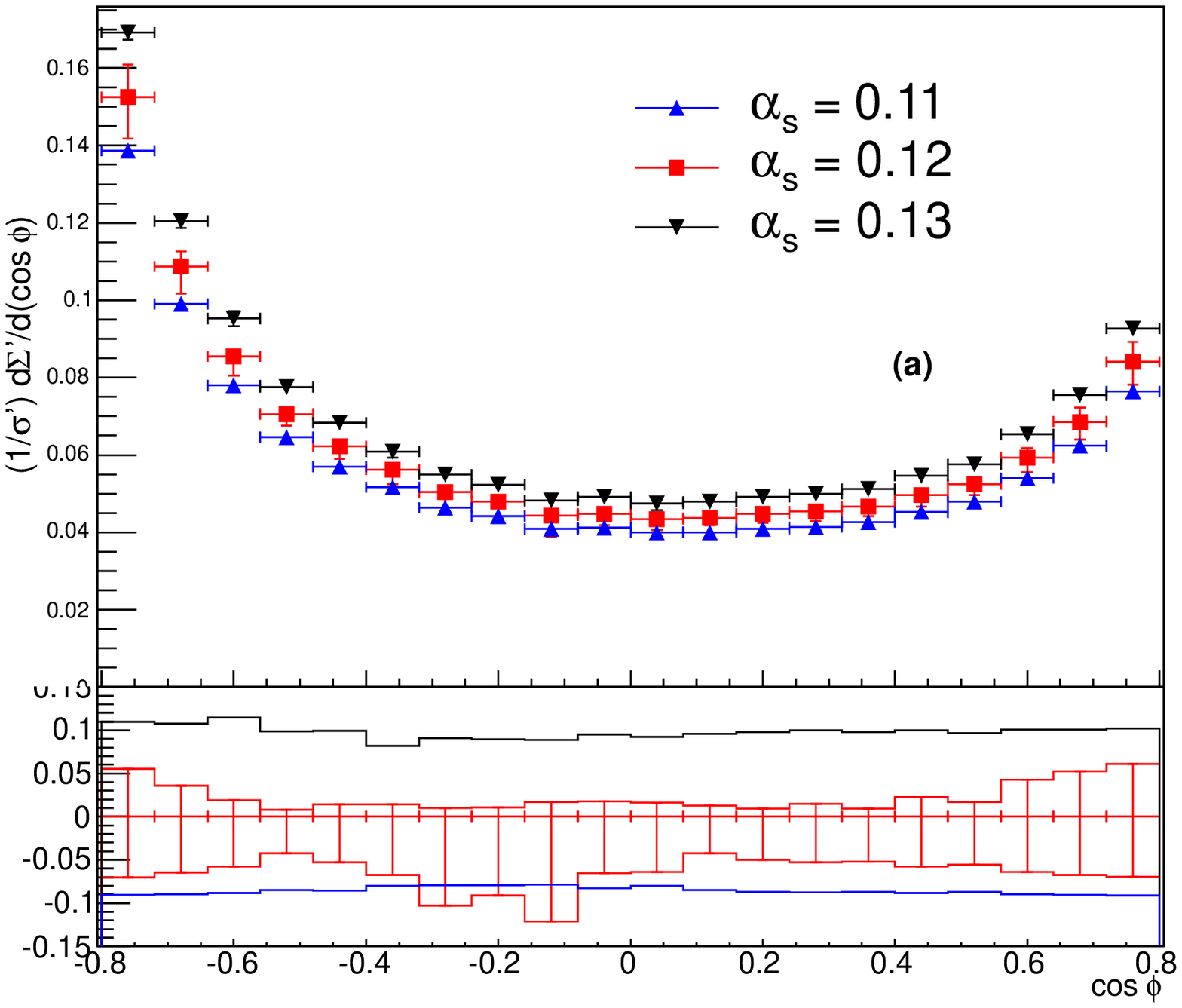}
\includegraphics[scale=0.44]{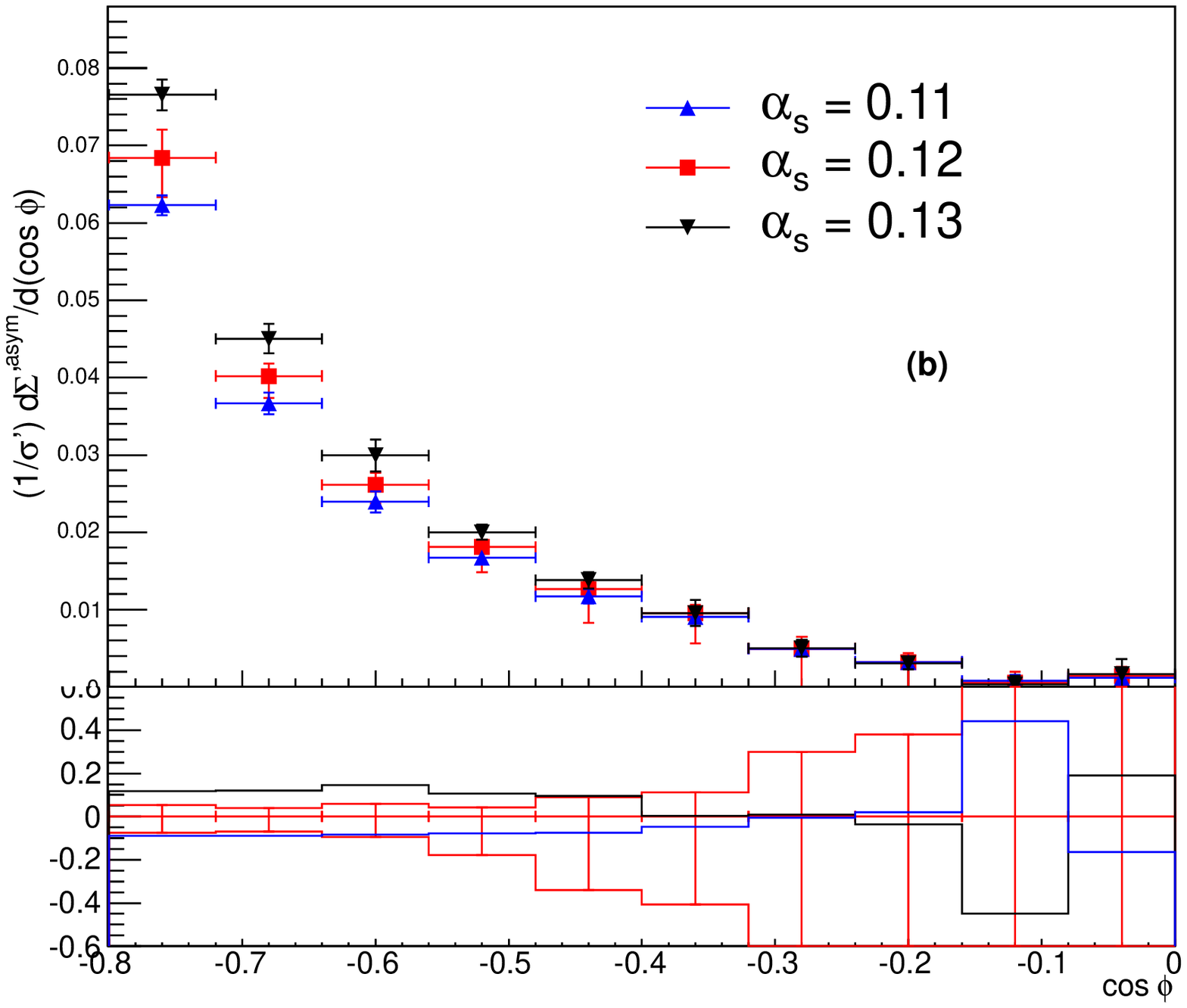}
\caption{(color online) Transverse EEC cross section (a) and its asymmetry (b) with three values 
 of $\alpha_s(M_Z)$  $=0.11$ (blue), $=0.12$ (red), and $= 0.13$ (black). The bottom panel of the figures demonstrate the size of errors (red) and deviations with  the values 
 of $\alpha_s(M_Z)$  $=0.11$ (blue) and $= 0.13$ (black) from the results evaluated with $\alpha_s=0.12$.   } \label{fig:as}
\end{center}
\end{figure}

%
Having detailed the intrinsic uncertainties from a number of dominant sources, we now wish to 
 investigate the sensitivity of the transverse
EEC and the AEEC on $\alpha_s(M_Z)$. In relating the strong coupling
 $\alpha_s(\mu)$ at a certain scale relevant for the collider jets, such as $\mu=E_T^{\rm max}$,
to the benchmark value $\alpha_s(M_Z)$, we have used the
two-loop $\beta$-function and the explicit formula for transcribing 
$\alpha_s(\mu)$ to $\alpha_s(M_Z)$ can be seen in Eq.~\eqref{eq:alpha_s}.
Results  for the transverse EEC and the AEEC are shown in  Fig.~\ref{fig:as} (a) and
Fig.~\ref{fig:as} (b), respectively, for the 
three indicated values  of $\alpha_s(M_Z)$:  $=0.11$(blue), $=0.12$ (red), $=0.13$ (black).
 The scale uncertainties are included only in the curve corresponding to
 $\alpha_s(M_Z)=0.12$, as it is close to the current world average
 $\alpha_s(M_Z)=0.1184$~\cite{Beringer:1900zz} and hence our focus on this value.
To demonstrate the intrinsic errors in the calculations of the transverse EEC and its 
asymmetry, we show the percentage size of the errors in 
the lower part of Fig.~\ref{fig:as} (a) and Fig.~\ref{fig:as} (b),
respectively, for $\alpha_s(M_Z)=0.12$.
 Concentrating first on the transverse EEC, we see that the bin-by-bin errors are
 typically $+2\%$ and $-6\%$ (for $|\cos \phi| \leq 0.6$), and somewhat larger
 for $|\cos \phi| > 0.6$.
A part of this error is of statistical origin in our Monte Carlo based theoretical
 calculations and
is reducible, in principle, with the help of a more effective importance sampling algorithm
in the event generation. However, a part of  the error is  irreducible, given the current
 theoretical (NLO) precision. This is quantified for the
 normalized integrated transverse EEC X-section over the $\cos \phi$ range shown in
the figures above, which largely removes the statistical (bin-by-bin) error:
\begin{equation}
\begin{array}{c|c|c|c}
 \alpha_s (m_Z)
&
0.11
&
0.12
& 
0.13
\\
\hline
\langle \frac{1}{\sigma^\prime}\frac{d\Sigma^\prime}{d\phi}\rangle 
&0.092^{+0.001}_{-0.005}  & 0.101^{+0.001}_{-0.005}  & 0.111^{+0.001}_{-0.005}
\end{array}.
\nonumber
\end{equation}

The computational error on the transverse AEEC is larger, as
shown in  Fig.~\ref{fig:as} (b) for $\alpha_s(M_Z)=0.12$. In particular, the errors for
the last four bins in the AEEC X-section are large due to the intrinsically small value
of this cross-section as $\cos \phi \to 0$. However, in the region
$-0.8 \leq \cos \phi \leq -0.4$, a clear dependence of
the differential transverse AEEC on  $\alpha_s(M_Z)$ is discernible. This is also displayed for
the normalized integrated transverse AEEC X-section given below (in units of $10^{-3}$),
in which the last four bins contribute very little:
\begin{equation}
\begin{array}{c|c|c|c}
 \alpha_s (m_Z)
&
0.11
&
0.12
& 
0.13
\\
\hline
\langle \frac{1}{\sigma^{\prime}}\frac{d\Sigma^{\prime~asymm}}{d\phi}\rangle 
&13.6^{+0.2}_{-1.4}  & 14.8^{+0.3}_{-1.5}  & 16.4^{+0.4}_{-1.6}
\end{array}.
\nonumber
\end{equation}

Details of the calculations and numerical results
for other values of the parameter $R$, cuts on $p_{\rm T min}$,  and the center-of-mass energies for the LHC and
the Tevatron will be published elsewhere.


\section{Summary} \label{sec:summary}

To summarize, we have  presented for the first time NLO results for the transverse EEC and
its asymmetry for jets at the LHC.  These distributions are shown  to 
have all the properties that are
required for the precision tests of perturbative QCD. In particular, they
(i) are almost independent of the structure functions, with typical uncertainties at $1\%$,
(ii) show weak scale sensitivity; varying the scale from $\mu=E_T/2$ to $\mu=2 E_T$, the
uncertainties are less than $5\%$ with the current (NLO) theoretical accuracy,
(iii) their dependence on modeling the underlying minimum bias events for judicious choice
of the parameter $R$ is likewise mild, ranging typically from $2\%$ to $5\%$
as one varies from $R=0.4$ to $R=0.6$, 
and (iv) preserve sensitivity to $\alpha_s(M_Z)$; varying $\alpha_s(M_Z)=0.11$ to 0.13, the
transverse EEC (AEEC) cross  section changes approximately by $20\%$ ($15\%$), and  
thus these distributions  will prove to be powerful techniques for the quantitative study
of event shape variables  and in the measurement of $\alpha_s(M_Z)$ in hadron colliders.

\section*{Acknowledgement}
We thank Zoltan Nagy for providing us his NLOJET++ code and helpful discussions. We also thank
Jan Kotanski for his help in the implementation of this program. Communications with
Gavin Salam and Markus Wobisch are also thankfully acknowledged. W. W. is supported by
the Alexander-von-Humboldt Stiftung.


\begin{thebibliography}{99}
%

\bibitem{Nakamura:2010zzi}
  K.~Nakamura {\it et al.}  [Particle Data Group],
  J.\ Phys.\ G {\bf 37}, 075021 (2010).

\bibitem{Wobisch:2012iu}
  For a review of recent QCD results from the Tevatron, see, for example, M.~Wobisch,
  Nucl. Phys. B Proc. Suppl. (2012) (in press) and arXiv:1202.0205 [hep-ex].

\bibitem{Chatrchyan:2012pb}
  S.~Chatrchyan {\it et al.}  [CMS Collaboration],
  arXiv:1204.0696 [hep-ex].

\bibitem{Aad:2010ad}
  G.~Aad {\it et al.}  [Atlas Collaboration],
  Eur.\ Phys.\ J.\  C {\bf 71}, 1512 (2011)
  [arXiv:1009.5908 [hep-ex]].

\bibitem{Catani:1993hr}
  S.~Catani, Y.~L.~Dokshitzer, M.~H.~Seymour and B.~R.~Webber,
  Nucl.\ Phys.\  B {\bf 406}, 187 (1993).

\bibitem{Catani:1996vz}
  S.~Catani and M.~H.~Seymour,
  Nucl.\ Phys.\  B {\bf 485}, 291 (1997)
  [Erratum-ibid.\  B {\bf 510}, 503 (1998)];
  [arXiv:hep-ph/9605323].

\bibitem{Cacciari:2008gp}
  M.~Cacciari, G.~P.~Salam and G.~Soyez,
  JHEP {\bf 0804}, 063 (2008)
  [arXiv:0802.1189 [hep-ph]].

\bibitem{Catani:1996jh}
  S.~Catani and M.~H.~Seymour,
  Phys.\ Lett.\  B {\bf 378}, 287 (1996)
  [arXiv:hep-ph/9602277].


\bibitem{Nagy:2001fj}
  Z.~Nagy,
  Phys.\ Rev.\ Lett.\  {\bf 88}, 122003 (2002)
  [arXiv:hep-ph/0110315];
  Phys.\ Rev.\  D {\bf 68}, 094002 (2003)
  [arXiv:hep-ph/0307268].

\bibitem{Banfi:2004nk}
  A.~Banfi, G.~P.~Salam and G.~Zanderighi,
  JHEP {\bf 0408}, 062 (2004)
  [arXiv:hep-ph/0407287];
  JHEP {\bf 1006}, 038 (2010)
  [arXiv:1001.4082 [hep-ph]].

\bibitem{Aaltonen:2011et}
  T.~Aaltonen {\it et al.}  [CDF Collaboration],
  Phys.\ Rev.\  D {\bf 83}, 112007 (2011)
  [arXiv:1103.5143 [hep-ex]].

\bibitem{Khachatryan:2011dx}
  V.~Khachatryan {\it et al.}  [CMS Collaboration],
  Phys.\ Lett.\  B {\bf 699}, 48 (2011)
  [arXiv:1102.0068 [hep-ex]].

\bibitem{Aad:2012np} 
  G.~Aad {\it et al.}  [ATLAS Collaboration],
  arXiv:1206.2135 [hep-ex].

\bibitem{Sjostrand:2006za}
  T.~Sjostrand, S.~Mrenna and P.~Z.~Skands,
  JHEP {\bf 0605}, 026 (2006)
  [arXiv:hep-ph/0603175].

\bibitem{Sjostrand:2007gs}
  T.~Sjostrand, S.~Mrenna and P.~Z.~Skands,
  Comput.\ Phys.\ Commun.\  {\bf 178}, 852 (2008)
  [arXiv:0710.3820 [hep-ph]].

\bibitem{Bahr:2008pv}
  M.~Bahr {\it et al.},
  Eur.\ Phys.\ J.\  C {\bf 58}, 639 (2008)
  [arXiv:0803.0883 [hep-ph]];  arXiv:1102.1672[hep-ph]]. 

\bibitem{Ali:1984yp}
  A.~Ali, E.~Pietarinen and W.~J.~Stirling,
  Phys.\ Lett.\  B {\bf 141}, 447 (1984).


\bibitem{Basham:1978bw}
  C.~L.~Basham, L.~S.~Brown, S.~D.~Ellis and S.~T.~Love,
  Phys.\ Rev.\ Lett.\  {\bf 41}, 1585 (1978);
  Phys.\ Rev.\  D {\bf 19}, 2018 (1979).

\bibitem{Ali:1982ub}
  A.~Ali and F.~Barreiro,
  Phys.\ Lett.\  B {\bf 118}, 155 (1982);
  Nucl.\ Phys.\  B {\bf 236}, 269 (1984).

\bibitem{Richards:1983sr}
  D.~G.~Richards, W.~J.~Stirling and S.~D.~Ellis,
  Nucl.\ Phys.\  B {\bf 229}, 317 (1983).

\bibitem{Ali:2010tw}
  A.~Ali and G.~Kramer,
  Eur.\ Phys.\ J.\  {\bf H36}, 245 (2011)
  [arXiv:1012.2288 [hep-ph]].


\bibitem{Bern:2011ep} 
 See, for instance,  Z.~Bern {\it et al.},
  arXiv:1112.3940 [hep-ph].

\bibitem{Martin:2009iq}
  A.~D.~Martin, W.~J.~Stirling, R.~S.~Thorne and G.~Watt,
  Eur.\ Phys.\ J.\  C {\bf 63}, 189 (2009).

\bibitem{Lai:2010vv}
  H.~L.~Lai {\it et. al},
  Phys.\ Rev.\  D {\bf 82}, 074024 (2010).

\bibitem{Nason:2004rx}
  P.~Nason,
  JHEP {\bf 0411}, 040 (2004)
  [arXiv:hep-ph/0409146].

\bibitem{Frixione:2007vw}
  S.~Frixione, P.~Nason and C.~Oleari,
  JHEP {\bf 0711}, 070 (2007)
  [arXiv:0709.2092 [hep-ph]].

\bibitem{matching}
We are aware of the ongoing NLL/NLO matching project within MATCHBOX: 
  S.~Platzer and S.~Gieseke,
  arXiv:1109.6256 [hep-ph].
 Private communications with Judith Katzy, Jan Kotanski and Simon Plaetzer are acknowledged.  

\bibitem{Banfi:2004yd}
  A.~Banfi, G.~P.~Salam and G.~Zanderighi,
  JHEP {\bf 0503}, 073 (2005)
  [arXiv:hep-ph/0407286].

\bibitem{Beringer:1900zz} 
  J.~Beringer {\it et al.}  [Particle Data Group Collaboration],
  Phys.\ Rev.\ D {\bf 86}, 010001 (2012).

\end{thebibliography}
\end{document}